\documentclass[10pt,letterpaper]{article}
\usepackage[top=0.85in,left=2.75in,footskip=0.75in]{geometry}

\usepackage{changepage}

\usepackage[utf8]{inputenc}

\usepackage{textcomp,marvosym}

\usepackage{fixltx2e}

\usepackage{amsmath,amssymb,amsfonts}

\usepackage{cite}

\usepackage{nameref} 

\usepackage{microtype}
\DisableLigatures[f]{encoding = *, family = * }

\usepackage{rotating}

\usepackage{subfig}  
\usepackage{overpic}  

\raggedright
\usepackage[aboveskip=1pt,labelfont=bf,labelsep=period,justification=raggedright,singlelinecheck=off]{caption}

\makeatletter
\renewcommand{\@biblabel}[1]{\quad#1.}
\makeatother

\date{}

\newcommand{\dx}{\, \mathrm{d}\mathbf{x}}

\begin{document}
\vspace*{0.35in}

\begin{flushleft}
{\Large
\textbf\newline{Margination of white blood cells - a computational approach by a hydrodynamic phase field model}
}
\newline
\\
Wieland Marth\textsuperscript{1},
Axel Voigt\textsuperscript{1,2,*}
\\
\bf{1} Institute of Scientific Computing, TU Dresden, Germany
\\
\bf{2} Center for Systems Biology Dresden, Germany
\\

%
%
* E-mail: axel.voigt@tu-dresden.de
\end{flushleft}
\section*{Abstract}

We numerically investigate margination of white blood cells and demonstrate the dependency on a number of conditions including hematocrit, the deformability of the cells and the Reynolds number. A detailed mesoscopic hydrodynamic Helfrich-type model is derived, validated and used for the simulations to provides a quantitative description of the margination of white blood cells. Previous simulation results, obtained with less detailed models, could be confirmed, e.g. the largest probability of margination of white blood cells at an intermediate range of hematocrit values and a decreasing tendency with increasing deformability. The consideration of inertia effects, which become of relevance in small vessels, also shows a dependency and leads to less pronounced margination of white blood cells with increasing Reynolds number.

\section*{Author Summary}

White blood cells can only properly function if they adhere to the vessel walls. To facilitate the adhesion, white blood cells migrate toward the vessel walls in blood flow through a process called margination. We model and simulate the process and analyse dependencies on a number of conditions including hematocrit, deformability and Reynolds number. A detailed mesoscopic hydrodynamic Helfrich-type model is derived, validated, and used for the simulations. Besideds the confirmation of previous results, obtained with less detailed models, we show a less pronounced margination of white blood cells with increasing Reynolds number. Inertia effects, previously neglected, become of relevance in small vessels, where the cells may experience Reynolds numbers of order unity or higher. These results are also of relevence for the margination of other cell types.


\section*{Introduction}

Various experimental and simulation studies of flowing blood have shown that red blood cells (RBCs)
concentrate in the center of the blood vessel. This can be explained by a lift force, arising from cell-wall 
hydrodynamic interactions, the high deformability of RBCs and their nonspherical shapes, see e.g. \cite{Fedosovetal_PRL_2012,Fedosovetal_SM_2014}. 
The lift force results in a migration of RBCs towards the center of the vessel and a RBC free layer near 
the wall. Differences in size, shape, and deformability are assumed to lead to different lift forces and thus a separation of cells with different mechanical 
properties within the blood vessel. White blood cells (WBCs) have a near-spherical shape and are not 
very deformable and thus mechanically different from RBCs. The lift force on WBCs is expected to be 
much lower than that on RBCs, which suggests, that WBCs may get marginated 
to the RBC free layer near the wall. This effect requires the interaction of RBCs and WBCs and is of
utmost importance for the functioning of the immune system, which requires the adhesion of WBCs to the
vessel wall.

While in principle understood, detailed investigations show a non-trivial dependence of WBC margination
on various blood flow properties including hematocrit $H_t$, flow rate, vessel geometry, and RBC 
aggregation \cite{Pearsonetal_AJP_2000,Abbittetal_AJP_2003,Jainetal_PLOSONE_2009}, e.g. a 
pronounced margination within an intermediate range of $H_t \approx 0.2–0.3$, and reduced WBC margination 
for lower and higher $H_t$. Only recently, such behavior could be explained through simulation studies in 2D \cite{Fedosovetal_PRL_2012}. 
It is argued that for low $H_t$, WBC margination turns out to be weak due to a low concentration of RBCs and thus less interaction, while at 
high $H_t$ WBC margination is attenuated due to interactions of marginated WBCs with RBCs near a wall, 
which significantly limit the time WBCs spend near a wall. This argumentation is confirmed by 3D simulations in an idealized blood vessel 
\cite{Fedosovetal_SM_2014,Takeishi_PhysRep_2014}. We here do not add any new physical explanation, but confirm the simulation results 
by using a different, more realistic modeling approach, which is based on a Helfrich-type curvature-elastic model \cite{Helfrich_ZNF_1973} with 
various constraints concerning membrane inextensibility and area conservation, cell-cell interaction and incompressible fluid flow of the blood plasma.
This approach contains all relevant physical phenomena and only measurable parameters and thus allows for quantitative predictions. 

The paper is organized as follows. In Methods and Models we first review existing modeling approaches, discussing pros and cons, before we introduce in detail the 
Helfrich-type curvature-elastic model. The effect of various parameters on the margination of WBCs is discussed in Results and Discussion. The SI further provides
information on thermodynamic consistency of the model, a detailed numerical approach by adaptive finite elements, a parallelization concept and model validation 
and convergence studies.

\section*{Methods and models}

\subsection*{Previous Models}
Previous simulation studies, which have been performed to describe WBC margination, are based on strong assumptions. The first simulation approach \cite{Freund_PF_2007} assumes an incompressible Stokes flow, the cells are modeled with a linear elastic membrane and a global area constraint is enforced. A boundary integral formulation is used for numerical discretization. More recently, a particle-based Lagrangian approach was used \cite{Fedosovetal_PRL_2012, Fedosovetal_SM_2014}. Here, RBCs and WBCs are described by a network model, where the cells are represented through triangulated surfaces. Penalty terms are used to ensure global volume and global area conservation as well as local area conservation for each surface element. The approach thus guarantees inextensibility for sufficiently small surface elements. Each membrane point is connected to the fluid through viscous friction. The dynamics of the fluid flow is described by the smoothed dissipative particle dynamics (SDPD) method, an approximation for the Navier-Stokes equations which is only precise, if the particle density is large enough. Furthermore, the incompressibility of the fluid is not guaranteed a priori and has to be controlled. In \cite{Takeishi_PhysRep_2014} a finite element approach is used for the RBCs, which are modeled as biconcave capsules and a Lattice-Boltzmann method for the fluid flow. The problems are coupled through an immersed boundary method. 

Our approach will improve on these methods by considering RBCs and WBCs using a Helfrich-type curvature elastic model with an inextensibility constraint, which goes beyond the linear elastic membrane, network model or biconcave capsule approach considered previously. Furthermore we will consider the full Navier-Stokes equations to account also for inertia effects, as RBCs in small vessels may experience Reynolds numbers of order unity or higher, especially, if the vessels are constricted due to diseases such as thrombosis, e.g. \cite{Barketal_JBM_2010,Vennemannetal_EF_2007}.

\subsection*{Helfrich-type models}

Helfrich-type modeling approaches have been applied to understand the complex motions and shape changes RBCs undergo within a flow field, e.g. tank-treating (TT) and tumbling (TB) motion \cite{Fischeretal_Science_1978}. Within a low Reynolds number regime, the Stokes limit is valid and various numerical approaches have also been considered in this limit to analyze the TT and TB motion \cite{Krausetal_PRL_1996, Bibenetal_PRE_2003, Beaucourtetal_PRE_2004, Bibenetal_PRE_2005, Veerapanemietal_JCP_2009,Ghigliottietal_JFM_2010,Sohnetal_JCP_2010,Kimetal_JCP_2010,Zhaoetal_JFM_2011}.  All models consider a membrane free energy
\begin{eqnarray}
{\cal{E}} = \int_\Gamma \frac{1}{2}b_N (H - H_0)^2 \; d\Gamma \label{eq:Helfrich_sharp}
\end{eqnarray}
with membrane $\Gamma(t)$, mean curvature $H$, spontaneous curvature $H_0$, and normal bending rigidity $b_N$. Lagrange multipliers are used to enforce a global area constraint or the stronger inextensibility constraint. The jump condition for the fluid stress tensor $\mathbf{S} = \mathbf{S}_{0,i} = -p \mathbf{I} + \nu_{0,i} \mathbf{D}$, with pressure $p$, fluid viscosity $\nu_0$, cell viscosity $\nu_i$ and deformation tensor $\mathbf{D} = \nabla \mathbf{v} + (\nabla \mathbf{v})^T$, with velocity $\mathbf{v}$, along the membrane than reads
\begin{align}
\!\!\! [\mathbf{S} \cdot \mathbf{n}]_\Gamma \!\!&=\!\! \frac{\delta {\cal{E}}}{\delta \Gamma} + \lambda_{global} H \mathbf{n} &\qquad \mbox{global area constraint,} \\
\!\!\! [\mathbf{S} \cdot \mathbf{n}]_\Gamma \!\!&=\!\! \frac{\delta {\cal{E}}}{\delta \Gamma} + \lambda_{local} H \mathbf{n} + \nabla_\Gamma \lambda_{local} &\qquad \mbox{local inextensibility constraint}, \label{eq:inex_sharp}
\end{align}
with outer normal $\mathbf{n}$ and the surface gradient $\nabla_\Gamma = \mathbf{P} \nabla$, where $\mathbf{P} = \mathbf{I} - \mathbf{n} \otimes \mathbf{n}$ denotes the projection operator. The Lagrange multipliers $\lambda_{global}$ and $\lambda_{local}$ are functionals of the fluid velocity $\mathbf{v}$ and are obtained by requiring $\frac{d}{dt} \int_\Gamma \; d\Gamma = \int_\Gamma H \mathbf{v} \cdot \mathbf{n} \; d \Gamma = 0$ for the global area constraint and $\nabla_\Gamma \cdot \mathbf{v} = 0$ along $\Gamma(t)$ for the local inextensibility constraint. The jump condition for the velocity reads $[\mathbf{v}]_\Gamma = 0$.

The linearity of the Stokes problem allows for efficient decoupled algorithms to solve for the Lagrange multipliers \cite{Bibenetal_PRE_2003,Sohnetal_JCP_2010,Zhaoetal_JFM_2011}. However, considering blood flow in small blood vessels, the Reynolds number at the scale of the RBC, can be of order unity and the Stokes limit is at least questionable. Modeling approaches, which consider also inertia effects, have recently been introduced \cite{Laadharietal_PF_2012,Salacetal_JFM_2012,Alandetal_JCP_2014}. All have found that the classical TB behavior is no longer observed at moderate Reynolds numbers. As such, a suppression of TB motion could have far reaching consequences also for the interaction of RBCs and thus also the margination of WBCs. We will in the following consider the full Navier-Stokes equations, which read  inside and outside the cells
\begin{eqnarray}
\rho (\partial_t \mathbf{v} + \mathbf{v} \cdot \nabla \mathbf{v}) - \nabla \cdot \mathbf{S} \!\!&=&\!\!0 \\
\nabla \cdot \mathbf{v} \!\!&=&\!\! 0
\end{eqnarray}
with density $\rho = \rho_{0,i}$. The global area constraint can be treated explicitly, which was e.g. used by \cite{Bonitoetal_MMNP_2011} within a front tracking method, proposed by \cite{Duetal_DCDS_2007} and used by \cite{Marthetal_JMB_2013,Hausseretal_IJBMBS_2013} for a phase-field model and also considered in \cite{Salacetal_JCP_2011} for a level-set approach. The local inextensibility constraint is more delicate and leads to additional non-linearities, which are so far only considered within a level set approach in \cite{Salacetal_JCP_2011,Laadharietal_PF_2012} and within a phase field approximation in \cite{Alandetal_JCP_2014}. All these models consider only one cell and it will be the main modeling contribution to extent the described approaches in \cite{Alandetal_JCP_2014} to model also interactions of cells in an efficient way.   

\subsection*{Hydrodynamic phase field models}

The method introduces auxiliary phase fields $\phi_i$ that distinguish the inside and the outside of each cell $i = 1, \ldots, n$. The inside and the outside are separated from each other by a diffuse layer, which marks the membrane. The phase field variables are defined as
\begin{equation}
 \phi_i(t,\mathbf{x}):=\tanh\left(\frac{r_i(t,\mathbf{x})}{\sqrt{2}\epsilon}\right) \label{eq:phasefield}
\end{equation}
where $\epsilon$ characterizes the thickness of the diffuse interface and $r_i(t,\mathbf{x})$ denotes the signed-distance function between $\mathbf{x} \in \Omega$ and its nearest point on $\Gamma_i(t)$ the membrane of cell $i$. Depending on $r_i$ we label the inside with $\phi_i\approx 1$ and the outside with $\phi_i\approx -1$. $\Gamma_i(t)$ is then implicitly defined by the zero level set of $\phi_i$. The cell phase can thus be defined as $\phi_{cell} \approx 1$ with $\phi_{cell} = \max_{\mathbf{x} \in \Omega} (\phi_1, \ldots, \phi_n)$ and the fluid phase as $\phi_{0} = - \phi_{cell} \approx 1$.

The dynamics are now governed by equations that couple these phase fields to the actual physical degrees of freedom. We consider the diffuse nondimensional Helfrich energies
\begin{eqnarray}
{\cal{E}}_i(\phi_i) & = & \frac{1}{2 \text{Re} \text{Be}_i} \int_\Omega \frac{1}{\epsilon} \left(\epsilon \Delta\phi_i-\frac{1}{\epsilon}(\phi_i^2-1)(\phi_i+H_0)\right)^2\; d\Omega, \label{diffuseEnergy}
\end{eqnarray}
with Reynolds number $\text{Re} = \frac{\rho U L}{\nu_0}$ and bending capillary numbers $\text{Be}_i = \frac{4\sqrt{2}}{3} \frac{\nu_0 U L^2}{b_{N,i}}$, where $U$ denotes a characteristic velocity, $L$ a characteristic length and $b_{N,i}$ the bending rigidity of cell $i$. In \cite{Duetal_Nonl_2005} formal convergence for $\epsilon\to 0$ to the nondimensional form of the sharp interface energy in eq. (\ref{eq:Helfrich_sharp}) is shown. 

Instead of a direct extension of the models in \cite{Alandetal_JCP_2014} which considers an $L^2$-gradient flow for $\phi_i$ and enforces a volume and local or global area constraint by Lagrange multipliers, we here introduce a $H^{-1}$ gradient flow, which directly ensures volume conservation and treat the global area constraint using a penalty approach. Such an approach was already considered for one cell, but without flow interactions, in \cite{Campeloetal_EPJE_2007}. It has the advantage to keep the equations local. We introduce the nondimensional penalty energies
\begin{eqnarray}
{\cal{E}}_{i,area}(\phi_i) & = & \frac{c}{2 \text{Re} \text{Be}_i} (\mathcal{A}_{0,i}-\mathcal{A}(\phi_i))^2, \label{diffusepenaltyEnergy}
\end{eqnarray}
with penalty parameter $c$ and initial and desired area of cell $i$, $\mathcal{A}_{0,i}$ and $\mathcal{A}(\phi_i) = \int_\Omega \frac{\epsilon}{2}|\nabla\phi_i |^2+\frac{1}{4\epsilon}(\phi_i^2-1)^2 d \Omega$, respectively. The last term converges to $\frac{2\sqrt{2}}{3} \int_{\Gamma_i} d\Gamma$ if $\epsilon\to 0$, see \cite{Du_JCP_2006}.

In addition, we require an interaction energy ${\cal{E}}_{int}$, to be defined below, such that the overall energy can be written as
\begin{equation}
{\cal{E}}(\phi_1, \ldots, \phi_n) = \sum_{i=1}^n ({\cal{E}}_i(\phi_i) + {\cal{E}}_{i,area}(\phi_i)) + {\cal{E}}_{int}(\phi_1, \ldots, \phi_n))
\end{equation}
and the evolution equations for $\phi_i$ read
\begin{align}
\partial_{t} \phi_i + \mathbf{v}\cdot \nabla \phi_i  & = \gamma\Delta\phi^\natural_i 
\label{eq:phi_t}
\end{align}
with a small positive mobility coefficient $\gamma$ and the nondimensional chemical potentials
\begin{align}
\phi^\natural_i & = \frac{\delta \mathcal{E}(\phi_1,\ldots,\phi_n)}{\delta \phi_i} = \frac{\delta \mathcal{E}_{i}(\phi_i)}{\delta \phi_i} + \frac{\delta \mathcal{E}_{i,area}(\phi_i)}{\delta \phi_i} + \frac{\delta \mathcal{E}_{int}(\phi_1, \ldots, \phi_n)}{\delta \phi_i}.  
\label{eq_mu}
\end{align}
We obtain
\begin{align}
\frac{\delta \mathcal{E}_{i}}{\delta \phi_i} = \frac{1}{ \text{Re}\text{Be}_i}\psi_i, \quad \!\!\!
\psi_i = \Delta \mu_i-\frac{1}{{\epsilon}^2}(3\phi_i^2+2H_0 \phi -1) \mu_i, \quad \!\!\!
\mu_i  = \epsilon \Delta\phi_i-\frac{1}{{\epsilon}}(\phi_i^2-1)(\phi_i+H_0).
\end{align}
The penalty terms read 
\begin{align}
\frac{\delta \mathcal{E}_{i,area}}{\delta \phi_i} = \frac{c_i}{\text{Re}\text{Be}_i} \kappa_i (\mathcal{A}_{0,i}-\mathcal{A}(\phi_i)), \quad \kappa_i = \epsilon \Delta\phi_i-\frac{1}{{\epsilon}}(\phi_i^2-1)\phi_i. 
\label{eq:penalty}
\end{align}

Now, we have to consider the interaction terms. Interaction in principle is computationally costly, as it turns the problem into a nonlocal one and requires the coupling of all phase field variables $\phi_1, \ldots, \phi_n$ and computations of the distance between cells. We here consider only steric interactions and model the short range repulsion by a Gaussian potential, which in the sharp interface description reads
\begin{equation}
{\cal{E}}_{i,int}(\Gamma_1,\ldots,\Gamma_n) = \sum_{j=1\atop j\neq i}^n \alpha \int_{\Gamma_i}  w_{j} d \Gamma, \quad \mbox{ with }w_j(\mathbf{x}) = \exp\left(-\frac{r_j^2(\mathbf{x})}{\epsilon^2}\right)
\end{equation}
where $w_j$ is an interaction function and describes the influence of cell $j$ on its environment. The interaction parameter $\alpha>0$ determines the strength of the repulsive interaction between cell $i$ and cell $j$ with respect to the evolution of cell $i$. Using eq. \ref{eq:phasefield}, the signed distance function $r_j$ can be computed within the diffuse interface region as
\begin{equation}
r_{j }= - \frac{\epsilon}{\sqrt{2}} \ln \frac{1+\phi_j}{1-\phi_j}  \quad \forall \mathbf{x}: |\phi_j(\mathbf{x})| < 1
\end{equation}
We thus can approximate the short range interaction function $w_j$ as
\begin{equation}
w_{j} \approx
\begin{cases}
  \exp \left(\frac{1}{2} (\ln \frac{1+\phi_j}{1-\phi_j})^2\right),  & \text{if } |\phi_j(\mathbf{x})|<1\\
  0 & \text{otherwise}
\end{cases}
\end{equation}
and consider the interaction potential within the phase-field description, which reads in nondimensional form
\begin{equation}
{\cal{E}}_{i,int} (\phi_1, \ldots, \phi_n) = \frac{1}{\text{Re}\text{In}} \int_{\Omega} B(\phi_i) \sum_{j=1\atop j\ne i}^n w_{j} d \Omega
\end{equation}
with $B(\phi_i) = \frac{1}{\epsilon} (\phi_i^2 - 1)^2$ being nonzero only within the diffuse interface around $\Gamma_i$ and $\text{In} = \frac{4 \sqrt{2}}{3} \frac{\nu_0 U}{\alpha}$ the interaction number. The interaction energy thus reads 
\begin{equation}
{\cal{E}}_{int}(\phi_1, \ldots, \phi_n) = \sum_{i=1}^n {\cal{E}}_{i,int}(\phi_1, \ldots, \phi_n)
\end{equation} 
and we obtain
\begin{align}
\frac{\delta {\cal{E}}_{int}(\phi_1, \ldots, \phi_n)}{\delta \phi_i} =  \frac{1}{\text{Re}\text{In} } (B^\prime(\phi_i) \sum_{j=1\atop j\ne i}^n w_j + w_i^\prime \sum_{j=1\atop j\ne i}^n B(\phi_j)).
\end{align}
As our approximation of $w_i$ is not differentiable, we define
\begin{equation}
w_{i}^\prime \approx
\begin{cases}
  \frac{-2\sqrt{2}}{\epsilon}\frac{r_i}{\phi_i^2-1} \exp\left(\frac{-r_i^2}{\epsilon^2}\right),  & \text{if } |\phi_i(\mathbf{x})|<1\\
  0 & \text{otherwise}.
\end{cases}
\end{equation}

\begin{figure}[!ht]
\centering
\includegraphics[width=0.7\textwidth]{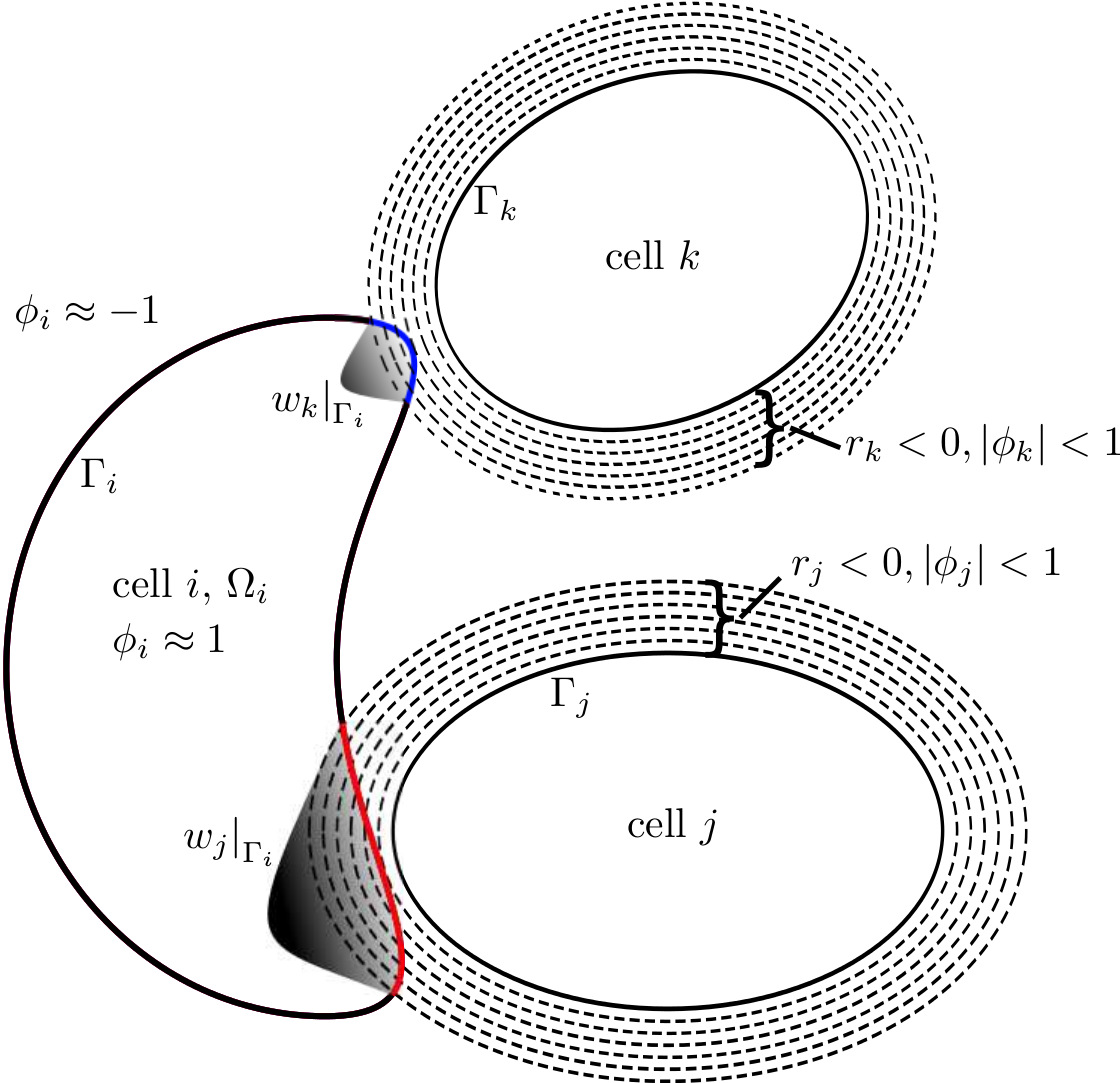}
\caption{The red and blue colored parts of $\Gamma_i$ are in contact with the interfaces of cell $j$ and cell $k$ (dashed contours around those cells), where the signed functions $r_j$ and $r_k$ can be calculated and thus also the interaction functions $w_j$ and $w_k$. They do not vanish in the overlapping regions.} \label{fig:sketch}
\end{figure}

Fig. \ref{fig:sketch} gives a schematic illustration of the interaction terms. The algorithm considers only these cells, for which the diffuse interfaces overlap. All other cells do not contribute to the interaction. In addition, the most expensive part, computing the distance between cells, has been avoided, as this information is already contained in the phase field description of the cells. The approach thus scales with $n$, the number of cells. Similar ideas to model interactions within phase field approaches have been considered in \cite{Zhangetal_JCP_2009,Guetal_JCP_2014}. However, only for the interaction of one cell with a fixed substrate.

The nondimensional Navier-Stokes equation reads
\begin{eqnarray}
\rho (\partial_{t} \mathbf{v} + \mathbf{v} \cdot \nabla \mathbf{v}) + \nabla p - \frac{1}{\text{Re}} \nabla \cdot (\nu \mathbf{D} ) \!\!&=&\!\! \sum_{i = 1}^n \phi_i^\sharp \nabla\phi_i \label{eq:navierStokesNonD1}\\
\nabla \cdot \mathbf{v} \!\!&=&\!\! 0 \label{eq:navierStokesNonD2}
\end{eqnarray}
with $\rho = 1$ and $\nu = \frac{1 - \phi_{cell}}{2} + \sum_{i=1}^n \frac{\nu_i}{\nu_0}\frac{\phi_i +1}{2}$. Different densities could be handled in a similar way but are omitted here for simplicity.

To enforce the local inextensibility constraint, we follow the approach of model B in \cite{Alandetal_JCP_2014}. The nonlinear evolution equations for $\phi_i$ remain, only the Navier-Stokes equation has to be extended and now reads
\begin{eqnarray}
\rho (\partial_{t} \mathbf{v} + \mathbf{v} \cdot \nabla \mathbf{v}) + \nabla p - \frac{1}{\text{Re}} \nabla \cdot (\nu \mathbf{D} ) \!\!&=&\!\! \sum_{i = 1}^n \phi_i^\sharp \nabla\phi_i + \nabla \cdot (\frac{|\nabla \phi_{cell}|}{2} \mathbf{P} \lambda_{local})  \label{eq:navierStokesNonD1_inex}\\
\nabla \cdot \mathbf{v} \!\!&=&\!\! 0 \label{eq:navierStokesNonD2_inex}
\end{eqnarray}
with a Lagrange multiplier $\lambda_{local}$ for which we introduce the additional equation 
\begin{equation}
\xi \epsilon^2 \nabla \cdot (\phi_{cell}^2 \nabla \lambda_{local}) + \frac{|\nabla \phi_{cell}|}{2} \mathbf{P} : \nabla \mathbf{v} = 0
\label{eq_lambda}
\end{equation}
with $\xi > 0$ a parameter independent of $\epsilon$. For $\epsilon \to 0$ we obtain $\Delta \lambda_{local} = 0$ away from $\Gamma = \cup_{1=i}^n \Gamma_i$ and $\mathbf{P} : \nabla \mathbf{v} = \nabla_\Gamma \cdot \mathbf{v} = 0$ near $\Gamma$, which was shown in \cite{Alandetal_JCP_2014} for $n = 1$.

In SI Analysis we show thermodynamic consistency of the derived models eq. \ref{eq:phi_t} and \ref{eq_mu} for $i = 1, \ldots, n$ and eq. \ref{eq:navierStokesNonD1} and \ref{eq:navierStokesNonD2} or eq. \ref{eq:navierStokesNonD1_inex}, \ref{eq:navierStokesNonD2_inex} and \ref{eq_lambda}. A detailed description of the numerical algorithm, which uses a finite element discretization in space and an operator splitting approach with a semi-implicit discretization in time is given in SI Numerics. The considered parallelization approach can be found in SI Implementation and numerical tests are provided in SI Benchmark Computations.

\section*{Results and Discussion}

 We study WBC margination for different WBC stiffnesses, different hematocrit values and different Reynolds numbers. We consider a blood vessel of thickness $20 \mu$m and length $40 \mu$m with periodic conditions on the in- and outflow. The relatively small length results from compromising computational efficiency and physical accuracy and has been obtained through detailed investigations on the influence of the periodicity on WBC margination. We consider RBCs with perimeter $22 \mu$m, area $19.5 \mu$m$^2$, bending rigidity $b_{N,RBC}=2\cdot 10^{-19}$ J, viscosity $\nu_{RBC} = 1 \cdot 10^{-3}$ $Pa\, s$. WBCs are initially set to be circular with radius $5 \mu$m. They have a viscosity $\nu_{WBC}=50\cdot10^{-2}$ Pa$\ $s. In order to study the influence of the stiffness of the WBCs, we consider three types: soft WBCs with $b_{N,WBC}= 2\cdot 10^{-19}$ J, hard WBCs with $b_{N,WBC}=2\cdot 10^{-18}$ J and rigid WBCs. The last is considered using a fluid particle dynamics (FPD) approach for the WBCs, see \cite{Tanaka_PRL_2000}. Please note that the FPD approach does not solve the phase field eq. \ref{eq:phi_t}, it rather updates the cell's midpoint by the averaged velocity inside the cell. The interaction strength is constant between all cell types and reads $\alpha = 4.24\cdot 10^{-7}$ N/m. For the fluid phase, we consider the viscosity $\nu_0 = 1 \cdot 10^{-3}$ $Pa\, s$. We consider a constant flow rate, which is realized by applying a time-dependent force term $\mathbf{F}=(\frac{1}{\text{Fr}(t)},0)^\top$, where $Fr$ denotes the Froude number. If the current flow rate $Q(t)$ is lower or greater than the desired flow rate $Q_0$, we increase the force term by multiplying it with the ratio of $Q_0/Q_t$. The initial force term can be estimated from its Newtonian value: $1/\text{Fr}(t=0)=12 Q_0/(H_l^3\text{Re})$, where $H_l$ is the channel height. We pick $Q_0 = 15$ for all simulations, which implies an averaged velocity of $8.44\cdot 10^{-5}$ m/s. An overview of all used parameters is given in Tab. \ref{tab:mat_parameters}.

\begin{table}
\centering
\begin{tabular}{|l|l|l|}
\hline
Symbol & Description & Value \\
\hline \hline
$L$ & radius of a perimeter-equivalent circular cell & $5\cdot 10^{-6}$ m \\
$U$ & characteristic velocity& $2.25 \cdot 10^{-5}$ m/s \\
$\rho$ & fluid density & $10^3$ kg/m$^{3}$ \\
$\nu_{0}$ & dynamic viscosity of the fluid & $10^{-3}$ Pa$\,$s\\
$\nu_{RBC}$ & dynamic viscosity of the RBC & $10^{-3}$ Pa$\,$s\\
$\nu_{W\!BC}$ & dynamic viscosity of the WBC & $5\cdot 10^{-2}$ Pa$\,$s\\
$b_{N,RBC}$ & bending rigidity of the RBC & $ 2\cdot 10^{-19}$ J \\
$b_{N,W\!BC}$ & bending rigidity of the hard WBC & $ 2\cdot 10^{-18}$ J \\
$b_{N,W\!BC}$ & bending rigidity of the soft WBC & $ 2\cdot 10^{-19}$ J \\
$\epsilon$ & diffuse interface thickness & 0.04\\
$\gamma$ & regularization parameter &  $10^{-7}$\\
$\alpha$ & repulsion parameter & $8.44\cdot 10^{-4}$ N/m\\
\hline
\end{tabular} 
\caption{Mechanical and numerical parameters used in the simulations. Mechanical parameters correspond to the considered values in \cite{Fedosovetal_PRL_2012,Takeishi_PhysRep_2014}. }
\label{tab:mat_parameters}
\end{table}

In nondimensional units, the computational domain becomes $\Omega=[0,8]\times[0,4]$, with periodic boundary conditions in the $x_1$ direction. The WBC has the radius 1 and is put at (5,2). RBCs are placed randomly such that they do not overlap. The nondimensional numbers read Re$=1.125\cdot10^{-4}$, Be$_{RBC}=5.3$, Be$_{WBC}=0.53$ (hard), Be$_{WBC}=5.3$ (soft), In$=0.1$ and Fr$(t=0)=4\cdot10^{-5}$. 

\begin{figure}[ht!]
\center
\includegraphics[width=0.3\textwidth]{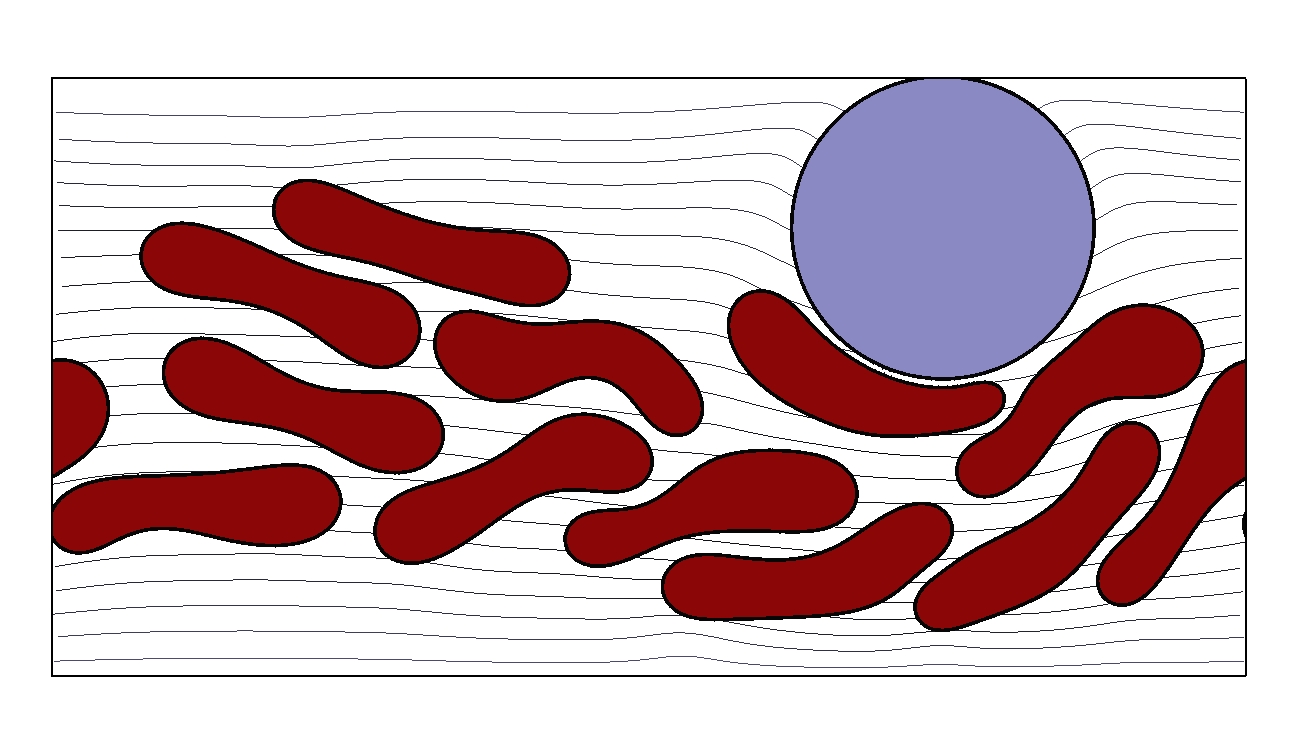}
\includegraphics[width=0.3\textwidth]{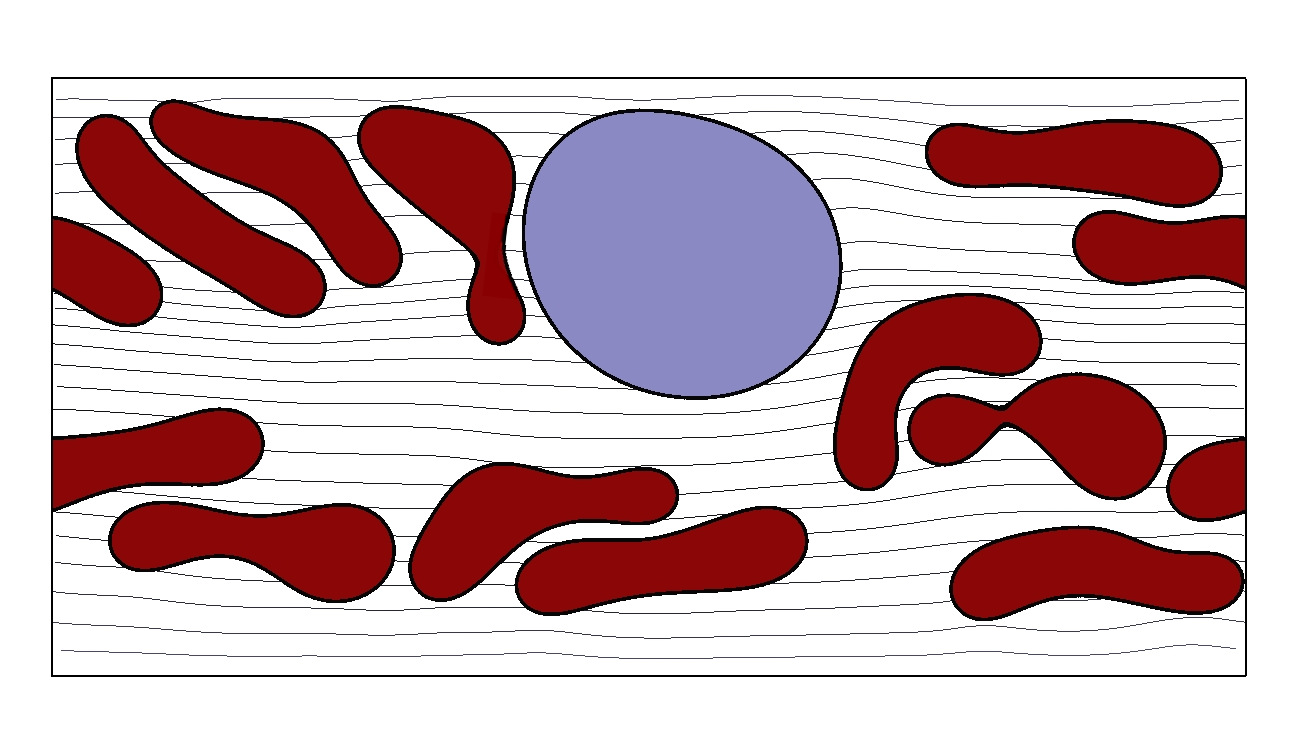}
\includegraphics[width=0.3\textwidth]{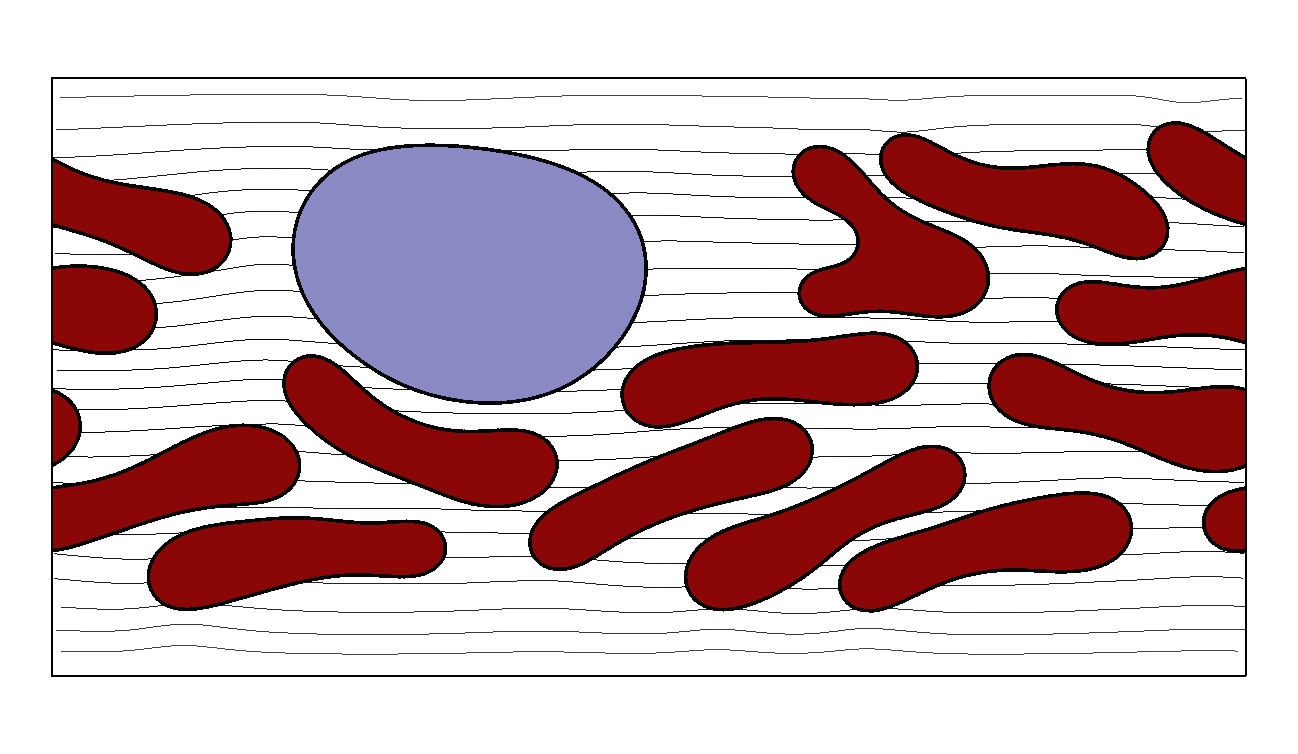} \\
\includegraphics[width=0.9\textwidth]{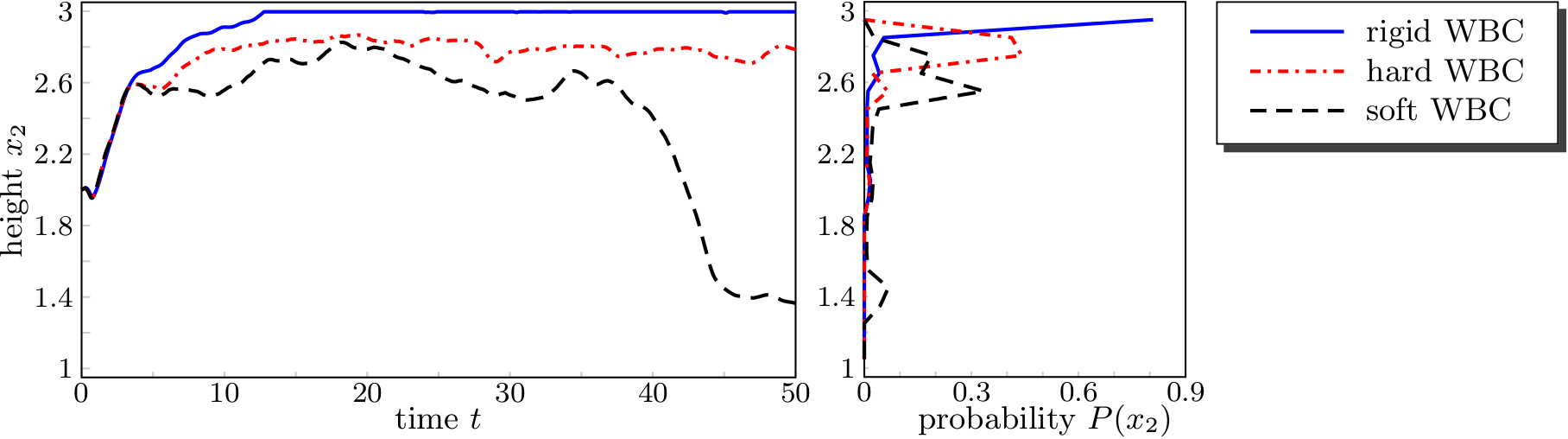}\label{fig:wbc_stiffness2}
\caption{Simulation snapshot at late time for $H_t=0.293$ for rigid, hard and soft WBC from left to right (top), $x_2$ coordinate for the trajectory of the midpoint of the WBC (bottom right) and the probability that the midpoint of the WBC is inside a defined interval (bottom left). $x_2$ axis is split into 20 intervals of length 0.1.}
\label{fig:wbc_stiffness}
\end{figure}

We first vary the deformability of the WBC and keep $H_t=0.293$ constant. The results are presented in Fig. \ref{fig:wbc_stiffness}, where the lower left diagram shows the $x_2$- coordinate of the trajectory of the midpoint of the WBC. After an initial phase, the WBC moves towards the wall, but only the rigid WBC can attach to the wall, while the soft WBC moves away after a certain time. The lower right diagram shows the probability that the midpoint of the cell is within the upper part of the channel with height $0.1$. The results nicely confirm the findings in 
\cite{Fedosovetal_PRL_2012}, that WBC margination is high for rigid cells and decreases for softer cells. 

\begin{figure}[ht!]
\center
\includegraphics[width=0.3\textwidth]{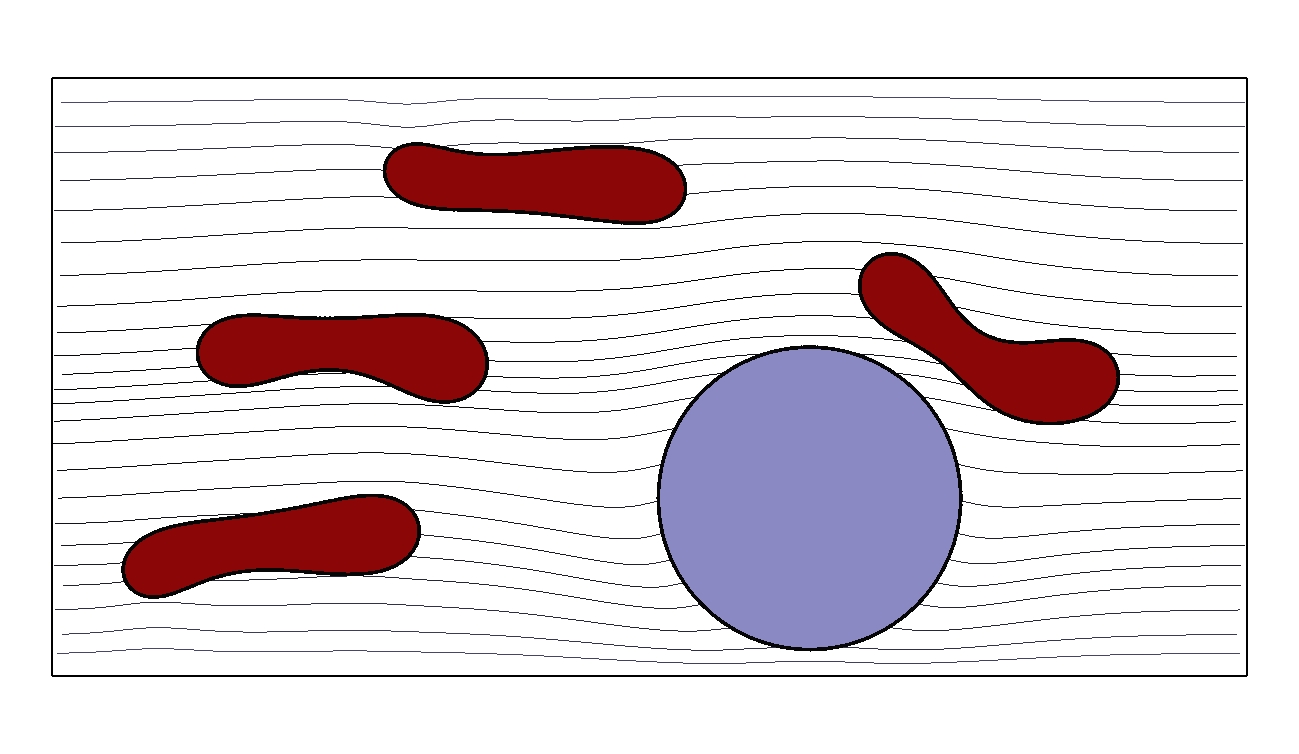}
\includegraphics[width=0.3\textwidth]{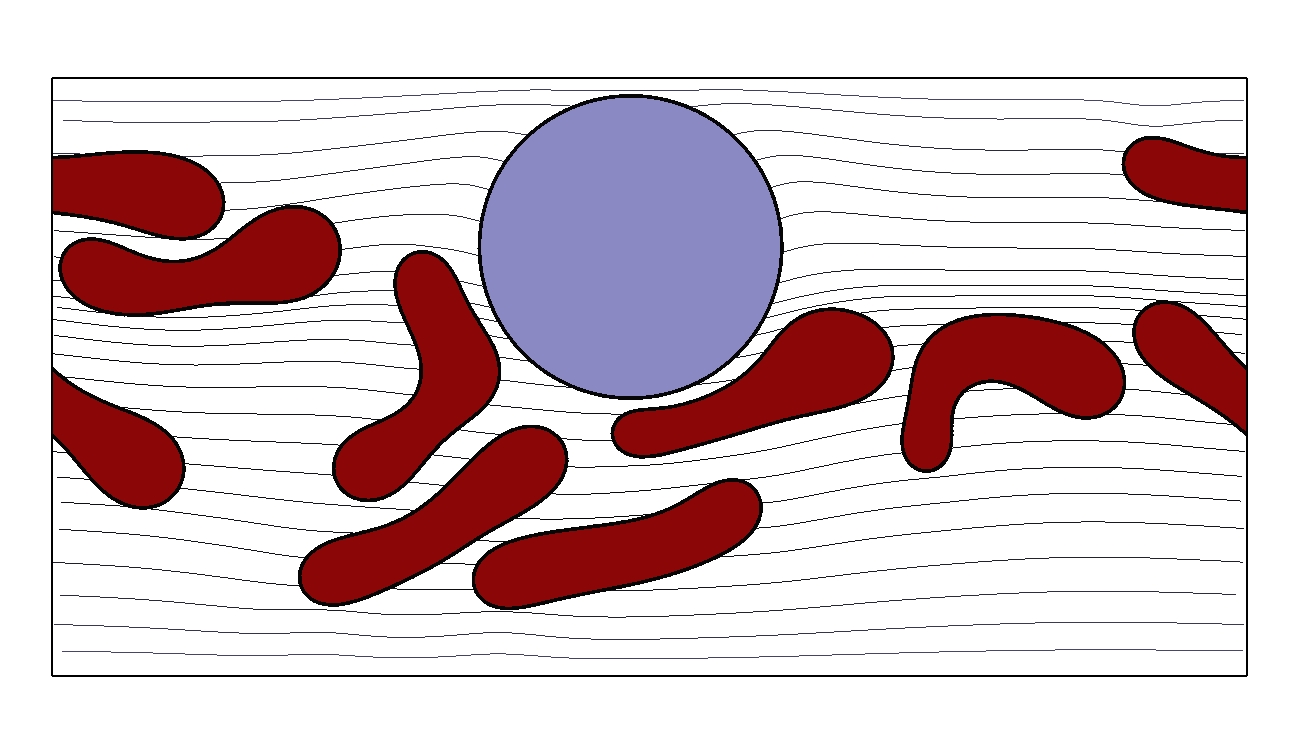}
\includegraphics[width=0.3\textwidth]{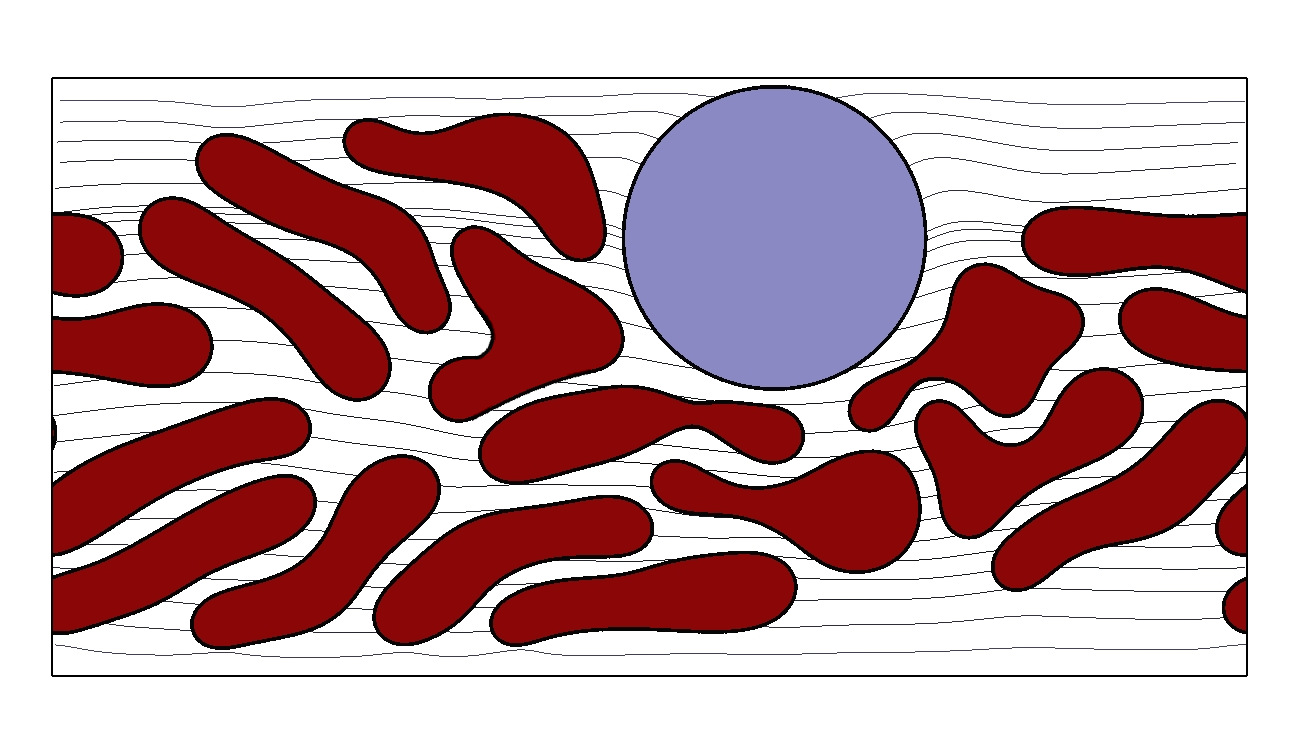} \\
\includegraphics[width=0.9\textwidth]{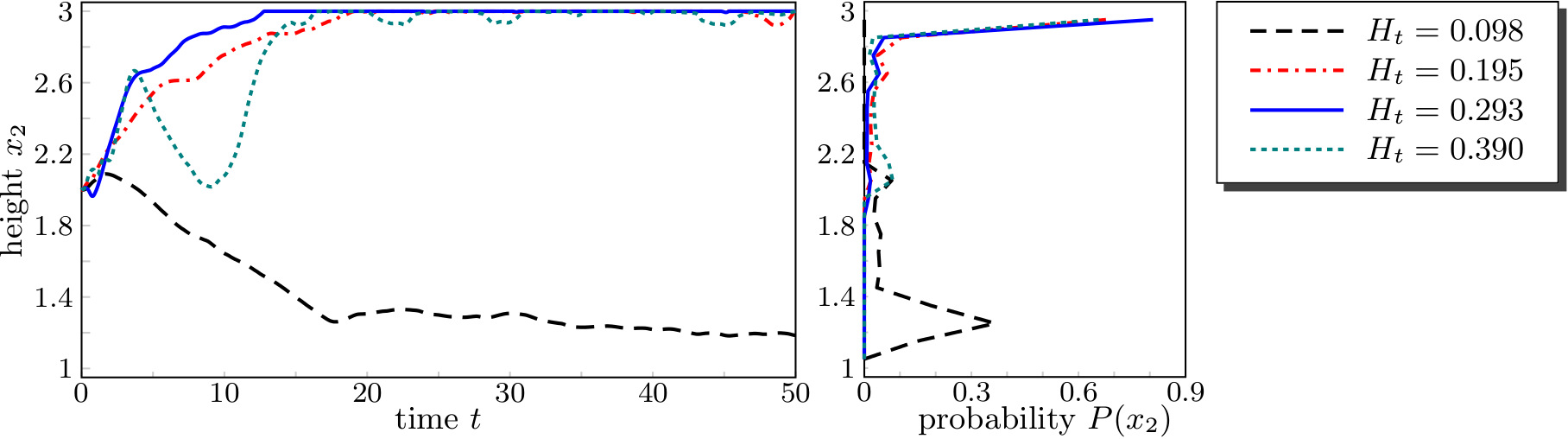}
\caption{Simulation snapshot at late time for a rigid WBC for $H_t=0.098$, $H_t=0.195$ and $H_t=0.39$ from left to right ($H_t=0.293$ is shown in Fig. \ref{fig:wbc_stiffness}) (top). $x_2$ coordinate for the trajectory of the midpoint of the WBC (bottom right) and the probability that the midpoint of the WBC is inside a defined interval (bottom left). $x_2$ axis is split into 20 intervals of length 0.1.}
\label{fig:wbc_rigid}
\end{figure}

\begin{figure}[ht!]
\center
\includegraphics[width=0.3\textwidth]{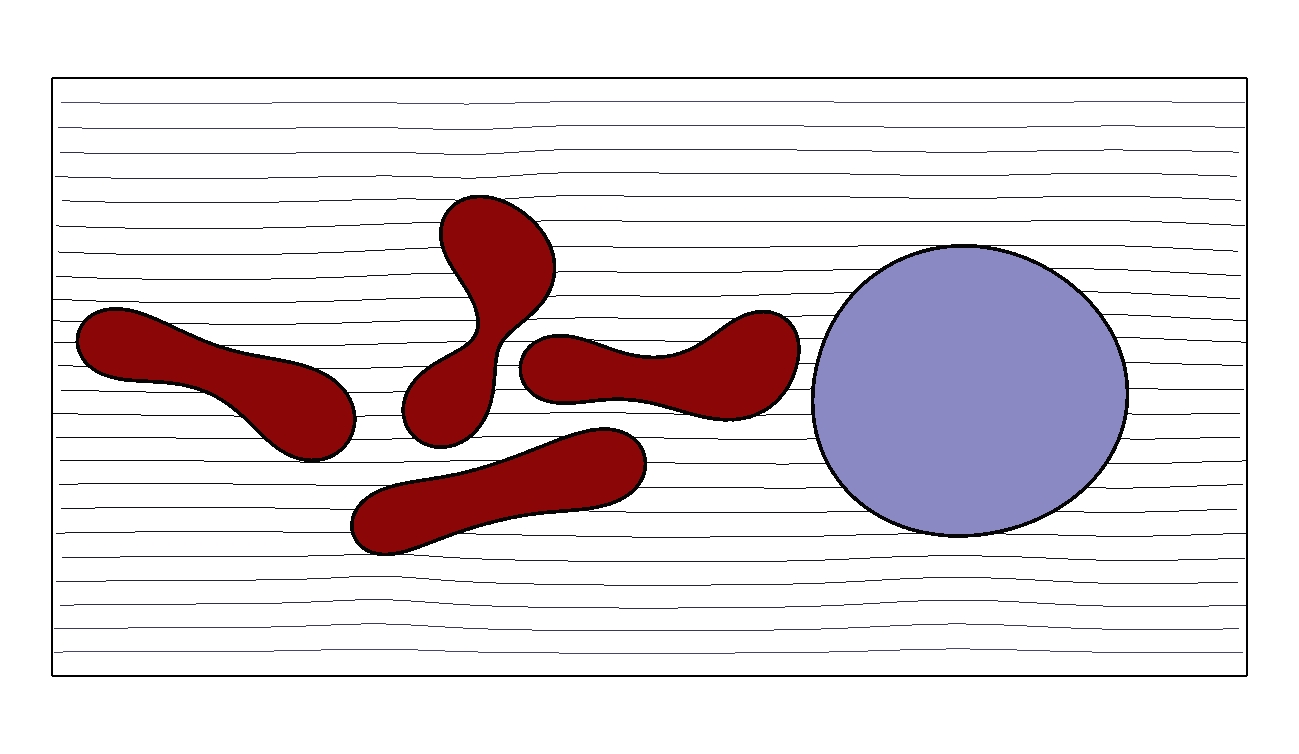}
\includegraphics[width=0.3\textwidth]{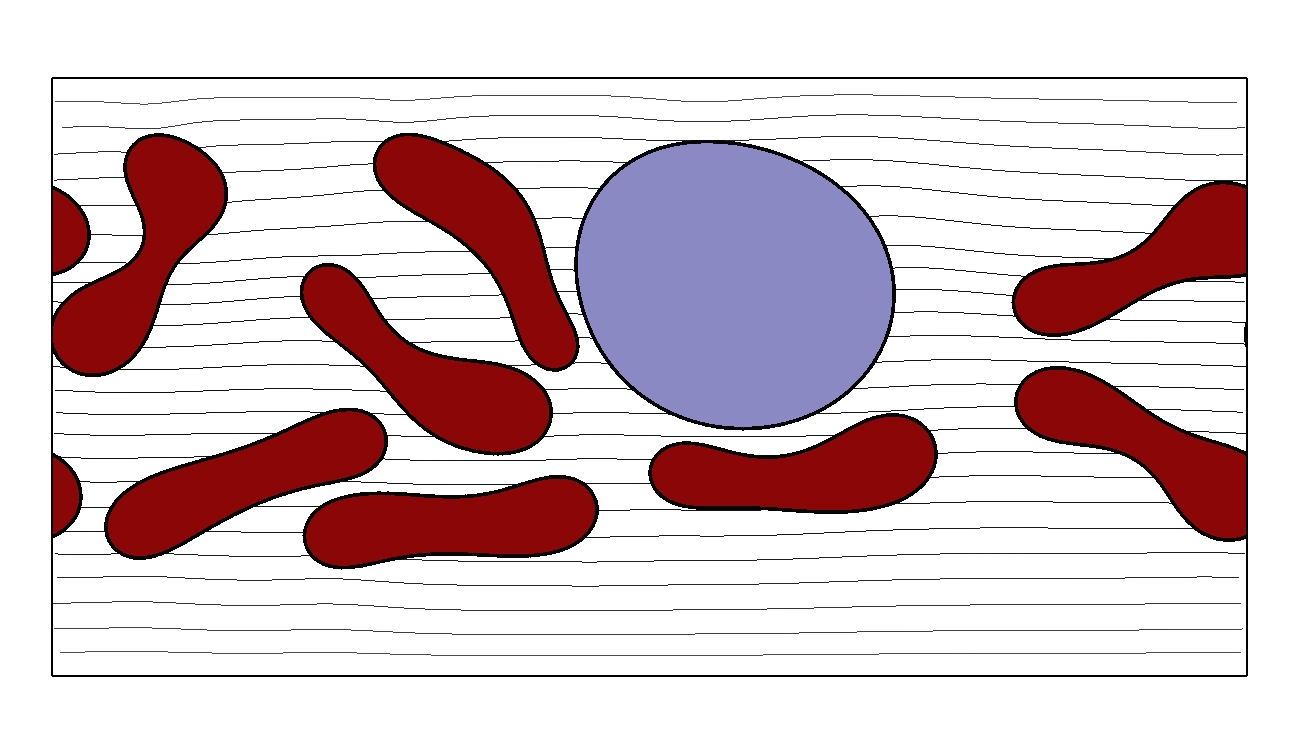}
\includegraphics[width=0.3\textwidth]{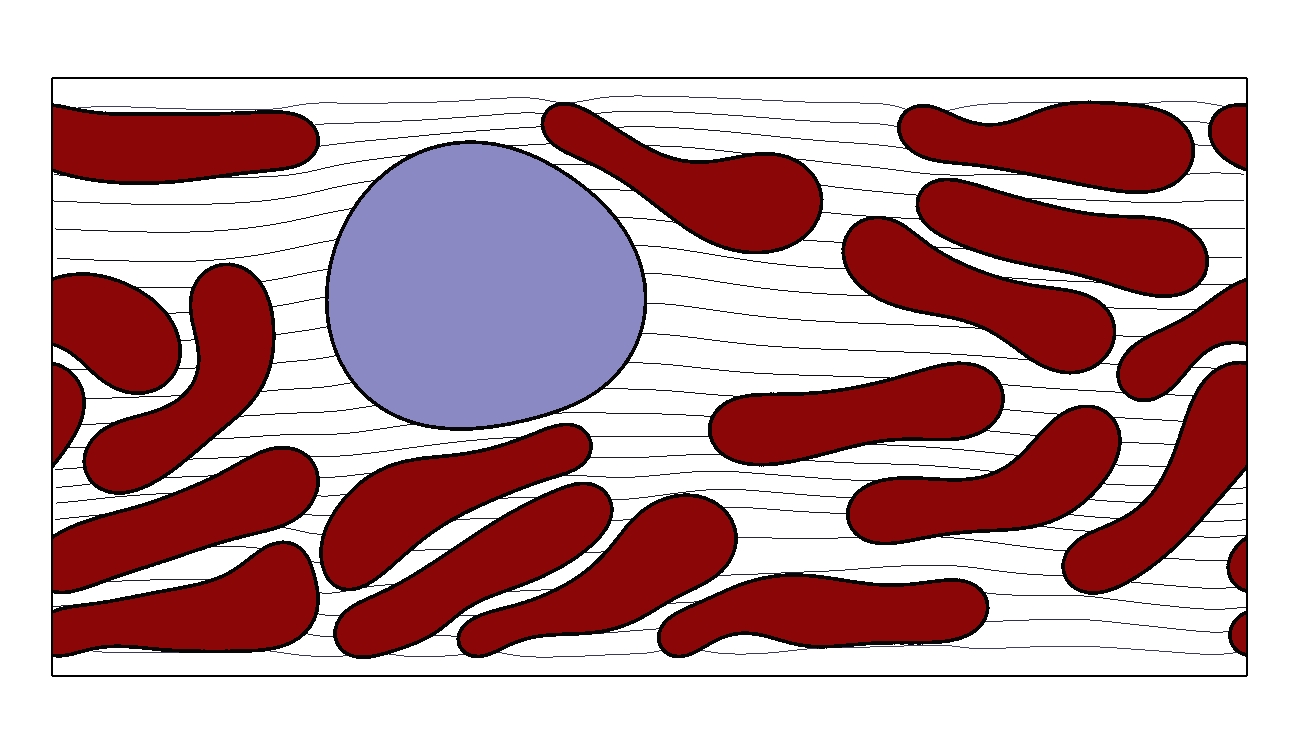} \\
\includegraphics[width=0.9\textwidth]{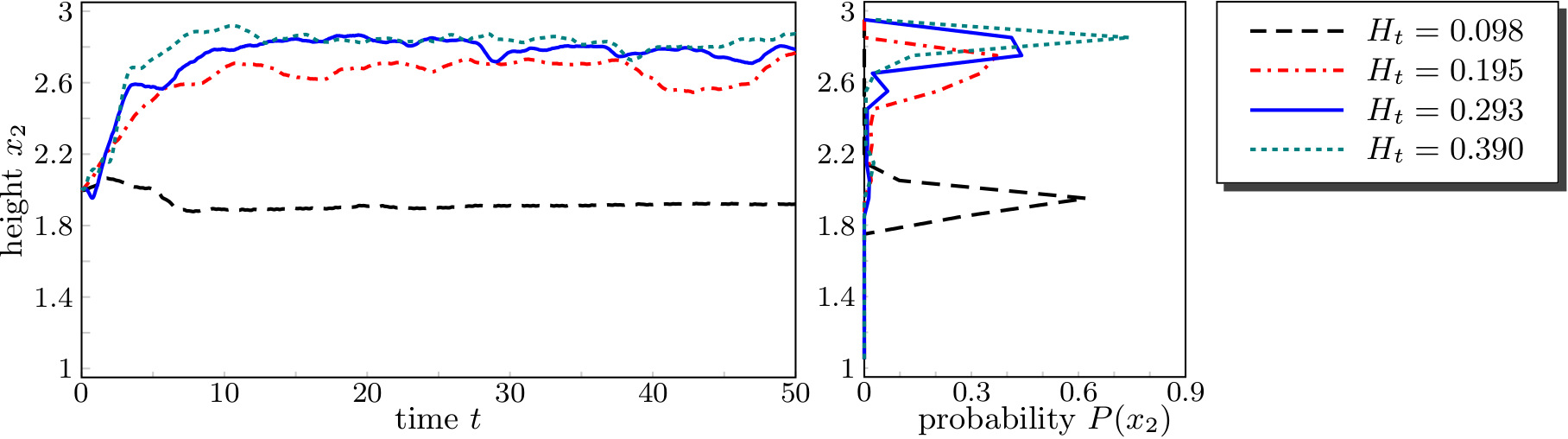}
\caption{Simulation snapshot at late time for a hard WBC for $H_t=0.098$, $H_t=0.195$ and $H_t=0.39$ from left to right ($H_t=0.293$ is shown in Fig. \ref{fig:wbc_stiffness}), (top). $x_2$ coordinate for the trajectory of the midpoint of the WBC (bottom right) and the probability that the midpoint of the WBC is inside a defined interval (bottom left). $x_2$ axis is split into 20 intervals of length 0.1.}
\label{fig:wbc_hard}
\end{figure}

The second test concerns the influence of $H_t$. We vary the number of RBCs, which lead to different values of $H_t$, ranging from 0.098 to 0.39. Fig. \ref{fig:wbc_rigid} shows the obtained results for a rigid WBC and Fig. \ref{fig:wbc_hard} for a hard one. 

For a rigid WBC, margination can be observed for all considered $H_t$. However, our simulations show a lower tendency to move to the wall for the smallest value of $H_t=0.098$ and the largest tendency for $H_t = 0.195$ and $H_t = 0.293$. For the highest value $H_t=0.39$, the probability slightly decreases. It seems more likely that due to the larger number of RBCs interaction between WBC and RBCs are possible also close to the wall, which moves the WBC away from the wall, see $t=10$, $t=22$ and $t=30$. This results give evidence for a decreasing WBC margination for high $H_t$, as also observed in \cite{Fedosovetal_PRL_2012}. In the case of a hard WBC, the cell remains in the center and in contrast to Fig. \ref{fig:wbc_rigid}, no margination occurs for the lowest $H_t$ considered. Increasing $H_t$ leads to WBC margination. However, contact with the wall cannot be achieved. We also do not see the tendency for decreasing WBC margination for $H_t = 0.39$. Further increasing $b_{N,WBC}$ or $H_t$ is not possible due to 
numerical reasons. 

\begin{figure}[ht!]
\center
\includegraphics[width=0.3\textwidth]{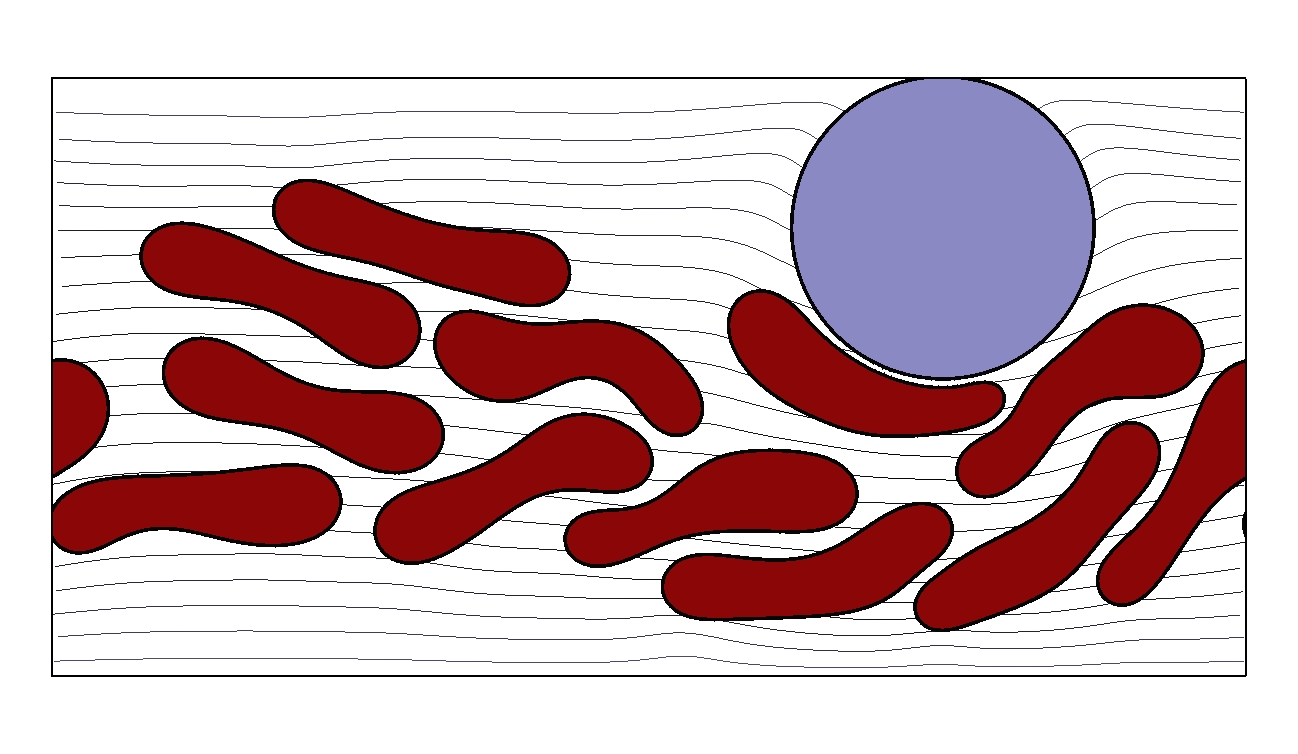}
\includegraphics[width=0.3\textwidth]{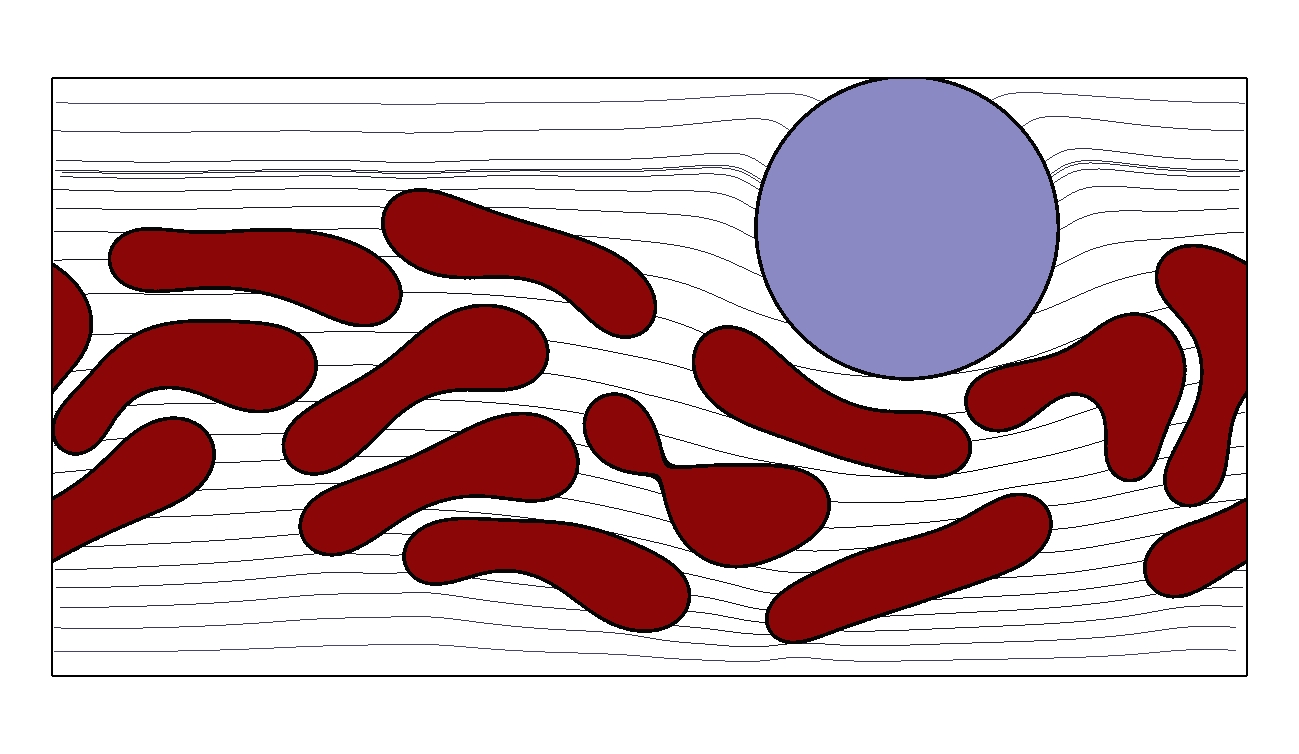}
\includegraphics[width=0.3\textwidth]{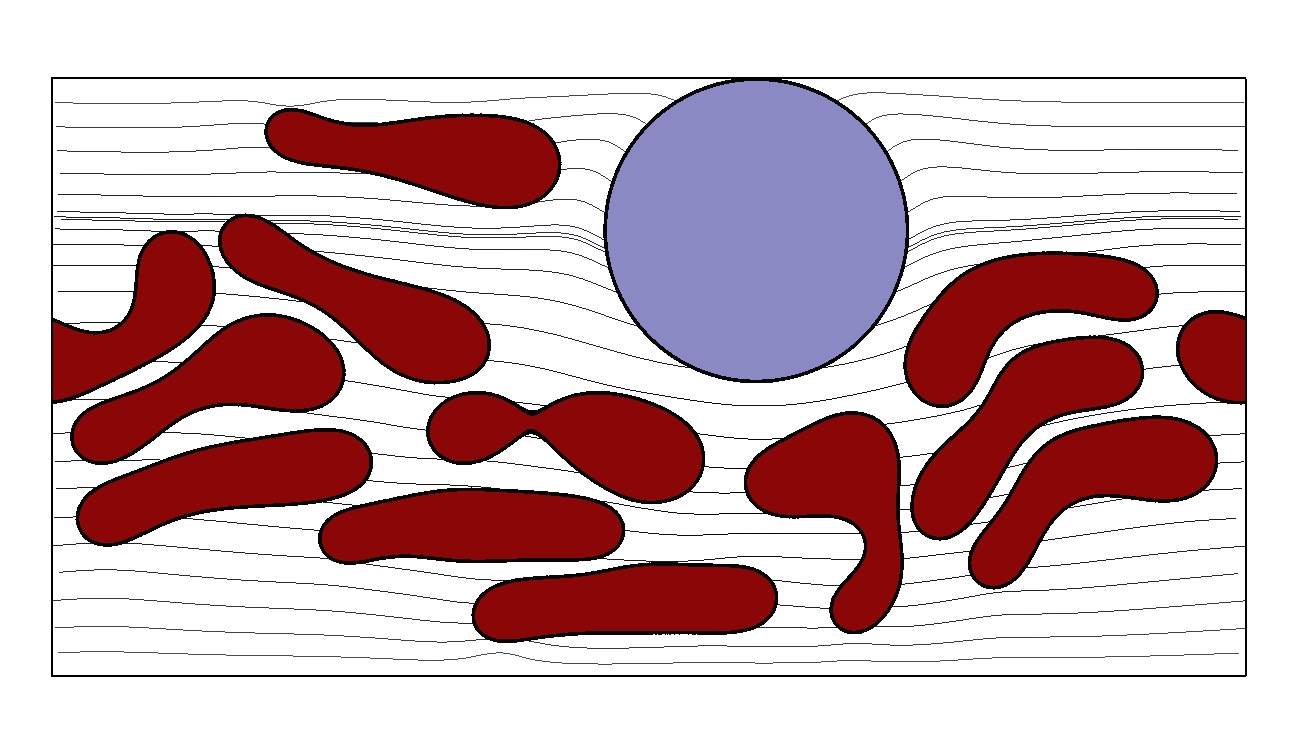} \\
\includegraphics[width=0.9\textwidth]{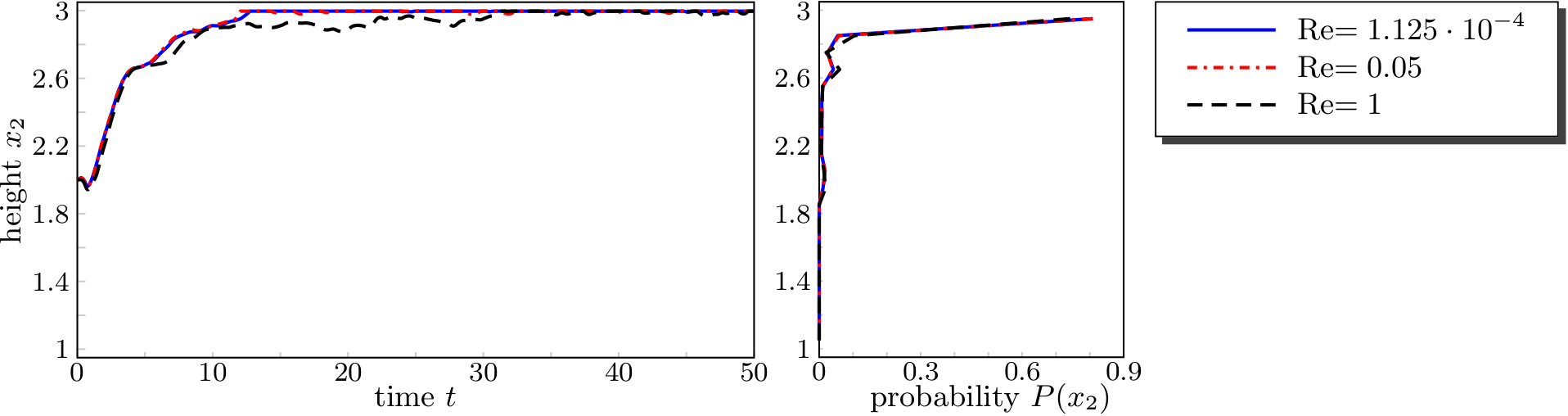}
\caption{Simulation snapshot at late time for a rigid WBC and $H_t = 0.293$ for Re=$1.125\cdot 10^{-4}$, Re=$0.05$ and Re=$1$ from left to right, (top). $x_2$ coordinate for the trajectory of the midpoint of the WBC (bottom right) and the probability that the midpoint of the WBC is inside a defined interval (bottom left). $x_2$ axis is split into 20 intervals of length 0.1.}
\label{fig:wbc_Re}
\end{figure}

The effect of the Reynolds number on WBC margination is shown in Fig. \ref{fig:wbc_Re}. We consider $H_t = 0.293$ and a rigid WBC. Considering a constant flow rate, we obtain WBC margination for Re=$1.125\cdot 10^{-4}$, Re=$0.05$ and Re=$1$. However, the tendency to adhere entirely decreases for higher Reynolds numbers. As Reynolds numbers of order unity are realistic for small vessels,  especially, if the vessels are constricted due to diseases, this effect might have severe influences on the functioning of the immune system. The simulation results for such Reynolds numbers indicate, that the tendency of RBCs to aggregate in the center of the vessel decreases. This increases the concentration of RBCs near the wall and thus leads to a stronger interaction with marginated WBCs, which, similar to the situation for large H$_t$, limits the time WBCs spend near the wall.

\section*{Supporting Information}

\subsection*{S1 Video}
\label{S1_Video}
{\bf Simulation video of WBC margination.} The video shows WBC margination in a vessel for $H_t=0.293$ considering a rigid WBC and adopting the simulation parameters Re$=1.125\cdot10^{-4}$, In$=0.1$ and $Q=15$.

\subsection*{S1 Analysis}
\label{S1_Text}
{\bf Thermodynamic consistency.} The proposed system of equations fulfill thermodynamic consistency. To show this, we consider the kinetic energy $\mathcal{E}_{kin}=\int \mathbf{v}^2$ with constant density $\rho=1$ and the cell energy $\mathcal{E}$ and show that the time derivative is less or equal to zero:
\begin{align}
\mathcal{\dot E}_{tot}(\mathbf{v},\phi_1, \ldots, \phi_n) & =\mathcal{\dot E}_{kin} + \mathcal{\dot E} = \int \mathbf{v} \mathbf{v}_t + \sum_{i=1}^n \phi_i^\natural \partial_t \phi_i d\mathbf{x}
\end{align}
with
\begin{align}
\partial_t \phi_i &= - \mathbf{v}\cdot \nabla \phi_i  + \gamma \Delta \phi_i^\natural\\
\partial_t \mathbf{v}  &= -(\mathbf{v}\cdot \nabla)\mathbf{v} - \nabla p +\frac{1}{\text{Re}}\nabla\cdot(\nu \mathbf{D})+\sum_{i = 1}^n \phi_i^\sharp \nabla\phi_i + \nabla \cdot (\frac{|\nabla \phi_{cell}|}{2} \mathbf{P} \lambda_{local})
\end{align}
which yields
\begin{align*}
\mathcal{\dot E}_{tot}(\mathbf{v},\phi_1, \ldots, \phi_n) & =  \int \mathbf{v} \cdot\left(-(\mathbf{v}\cdot \nabla)\mathbf{v} - \nabla p +\frac{1}{\text{Re}}\nabla\cdot(\nu \mathbf{D})+\sum_{i = 1}^n \phi_i^\sharp \nabla\phi_i\right. \\
&\quad \left.+ \nabla \cdot (\frac{|\nabla \phi_{cell}|}{2} \mathbf{P} \lambda_{local})\right)+ \sum_{i=1}^n \phi^\natural_i(- \mathbf{v}\cdot \nabla \phi_i  + \gamma \Delta \phi_i^\natural ) \dx\\
& \qquad \text{\footnotesize{\bigg(partial integration, use $\nabla \cdot \mathbf{v} = 0$\bigg)}}\\
& = \quad \int - \frac{1}{\text{Re}}|\nabla \mathbf{v}|^2 -\gamma \sum_{i=1}^n|\nabla \phi_i^\natural|^2  -(\nabla \mathbf{v}:\frac{|\nabla \phi_{cell}|}{2} \mathbf{P}) \lambda_{local}\dx \\
& \qquad \text{\footnotesize{\bigg(use eq. \ref{eq_lambda} and partial integration \bigg)}}\\& \quad  \\
& = \quad \int - \frac{1}{\text{Re}}|\nabla \mathbf{v}|^2 -\gamma \sum_{i=1}^n|\nabla \phi_i^\natural|^2  -\xi \epsilon^2 \phi^2_{cell} |\nabla\lambda_{local}|^2 \dx\\
& \leq 0,
\end{align*}
where we have used the identity $\mathbf{v} \times ( \nabla \times \mathbf{v}) = \nabla (|\mathbf{v} |^2) - (\mathbf{v} \cdot \nabla) \mathbf{v}$ from which follows that $\int \mathbf{v} \cdot (- \mathbf{v} \cdot \nabla \mathbf{v}) = 0$.

\subsection*{S2 Numerics}
\label{S2_Text}
{\bf Time discretization.} In order to discretize in time, we explore an operator splitting approach. In an iterative process, we first solve the flow problem and substitute its solution into the phase-field equations, which are then solved separately with a parallel splitting method. We split the time interval $I = [0,T]$ into equidistant time instants $0 = t_0 < t_1 < \ldots$ and define the time steps $\tau := t_{n+1}- t_n$. Of course, adaptive time steps may also be used. We define the discrete time derivative $d_t \cdot ^{n+1} := (\cdot ^{n+1} - \cdot ^{n})/ \tau$, where the upper index denotes the time step number and e.g $\mathbf{v}^n:=\mathbf{v}(t_n)$ is the value of $\mathbf{v}$ at time $t_n$. For each system, a semi-implicit time discretization is used, which together with an appropriate linearization of the involved non-linear terms leads to a set of linear system in each time step. \\

\vspace*{0.3cm}
\noindent {\bf Space discretization.} We apply the finite element method to discretize in space, where a $P^2/P^1$ Taylor-Hood element is used for the flow 
problem, all other quantities are discretized in space using $P^2$ elements.
In each time step we solve:
\begin{enumerate}
\item the flow problem for $\mathbf{v}^{n+1}$, $p^{n+1}$ and $\lambda_{local}^{n+1}$
\begin{align}
& d_t \mathbf{v}^{n+1} +(\mathbf{v}^n\cdot\nabla)\mathbf{v}^{n+1} = \\
& \quad  -\nabla p^{n+1} + \frac{1}{\text{Re}}\nabla\cdot(\nu^n\mathbf{D}^{n+1}) + \sum_{i=1}^n{\phi^\natural_i}^n \nabla\phi_i^n+\nabla\cdot(\frac{|\nabla\phi_{cell}^n|}{2}\mathbf{P}^n\lambda_{local}^{n+1}), \\
& \nabla \cdot \mathbf{v}^{n+1} =0 \\
& \xi \epsilon^2 \nabla \cdot ((\phi_{cell}^n)^2 \nabla \lambda_{local}^{n+1}) + \frac{|\nabla \phi_{cell}^{n}|}{2} \mathbf{P}^{n} : \nabla \mathbf{v}^{n+1} = 0
\end{align}
where $\nu^n=\nu(\phi^n)$ and $\mathbf{P}^n=\mathbf{I}-\frac{\nabla\phi^n\otimes\nabla\phi^n}{|\nabla\phi^n|^2}$
\item the phase field equations for $\phi_i^{n+1}$, $i=1,\ldots,n$
\begin{align}
d_t\phi_i^{n+1} + \mathbf{v}^{n+1}\cdot \nabla \phi_i^{n+1}  & = \gamma\Delta{\phi^\natural_i}^{n+1},\\
{\phi^\natural_i}^{n+1} &= \frac{1}{ \text{ReBe}_i}\psi_i^{n+1} + \frac{c}{\text{ReBe}_i}(\mathcal{A}_{0,i}-\mathcal{A}(\phi_i^n))\kappa_i^n\\
& \quad +\frac{1}{\text{ReIn}}\left( B^\prime(\phi_i^n)\sum_{j=1\atop j\ne i}^n w_j^n+{w_i^\prime}^n \sum_{j=1\atop j\ne i}^n B(\phi_j^n)\right),\\
\psi_i^{n+1} &= \Delta \mu_i^{n+1}-\frac{1}{\epsilon^2}(3({\phi_i^{n+1}})^2+2H_0\phi_i^{n+1}-1) \mu_i^{n+1},\\
\mu_i^{n+1}  &= \epsilon\Delta\phi_i^{n+1}-\frac{1}{\epsilon}(({\phi_i^{n+1}})^2-1)(\phi_i^{n+1}+H_0)\\
\end{align}
where we either choose $\kappa_i^n=-\epsilon\Delta\phi_i^{n}+\frac{1}{\epsilon}(({\phi_i^{n}})^2-1)\phi_i^{n}$ or $\kappa_i^n=\nabla\cdot(\frac{\nabla\phi_i^n}{|\nabla\phi_i^n|)})$. We again linearize the non-linear terms by a Taylor expansion of order one, e.g. $((\phi_i^{n+1})^2 -1)\phi_i^{n+1}=((\phi_i^n)^2-1)\phi_i^n+(3{(\phi_i^n)}^2-1)(\phi_i^{n+1}-\phi_i^n)$.

\end{enumerate}

\subsection*{S3 Implementation}
\label{S3_Text}
{\bf Implementation.} 
The fully discretized system of partial differential equations is implemented using the adaptive finite element toolbox AMDiS \cite{Veyetal_CVS_2007,Witkowskietal_ACM_2015}. We use an adaptively refined triangular mesh $\mathcal{T}_h$ with a high resolution along the cell membranes to guarantee at least five grid points across the diffuse interface. The criteria to refine or coarsen the mesh is purely geometric and related to the phase field variables $\phi_i$. Due to the use of adaptivity, we need to interpolate the old solution defined on ${\cal{T}}_h^n$ onto the new mesh ${\cal{T}}_h^{n+1}$. To do this without violating the conservation of cell volume, we solve $\langle \phi_i^{n,old}, \theta\rangle = \langle \phi_i^{n,new}, \theta \rangle$ in every adaption step, with test function $\theta$ and $\phi_i^{n,new}$ defined on ${\cal{T}}_h^{n+1}$ and $\phi_i^{n,old}$ on ${\cal{T}}_h^{n}$. We use a multi-mesh strategy \cite{Voigtetal_JCS_2012} to virtually integrate the first term on the finest common mesh ${\cal{T}}_h^{n} \cup {\cal{T}}_h^{n+1}$, which guarantees a constant cell volume as long as time steps are appropriately chosen. We require the cell membranes not to propagate over a whole element within one time step. With this restriction, all numerical tests show that $\int_\Omega \phi_i \; d \mathbf{x}$ is conserved. 
We conduct a shared memory OPENMP parallelization, to solve the phase field evolutions via a parallel splitting method. Furthermore, each linear system of equations is solved using the direct unsymmetric multifrontal method UMFPACK.

\subsection*{S4 Benchmark computations}
\label{S4_Text}
{\bf Benchmark - collision of two cells}
In \cite{Salacetal_JCP_2011, Alandetal_JCP_2014} it has been shown that two inextensible cells in an extensional flow do not coalesce, even without an interaction potential. If this would also be true in more general cases, it would allow to use one phase field variable for all cells and drastically simplify the considered model. We therefore consider the interaction of two cells in more detail and demonstrate that the considered model is indispensible.

\begin{figure}[ht!]
\begin{tabular}{c@{} c@{} c@{} c@{} c}
\vspace{-2.5ex}
\raisebox{5ex}{\footnotesize\bf{(a)}} & 
\subfloat{
\includegraphics[width=0.21\textwidth]{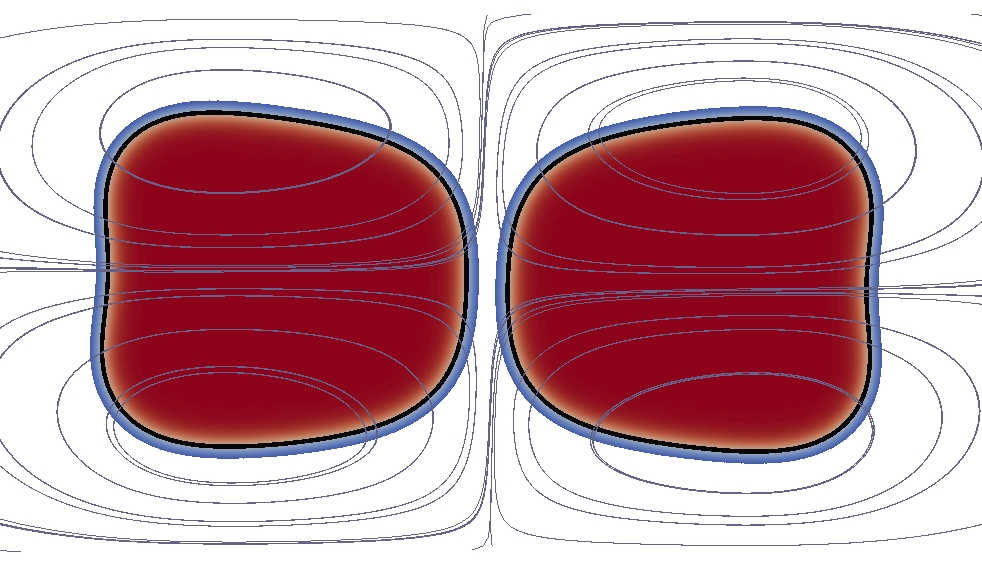}
}&
\subfloat{
\includegraphics[width=0.21\textwidth]{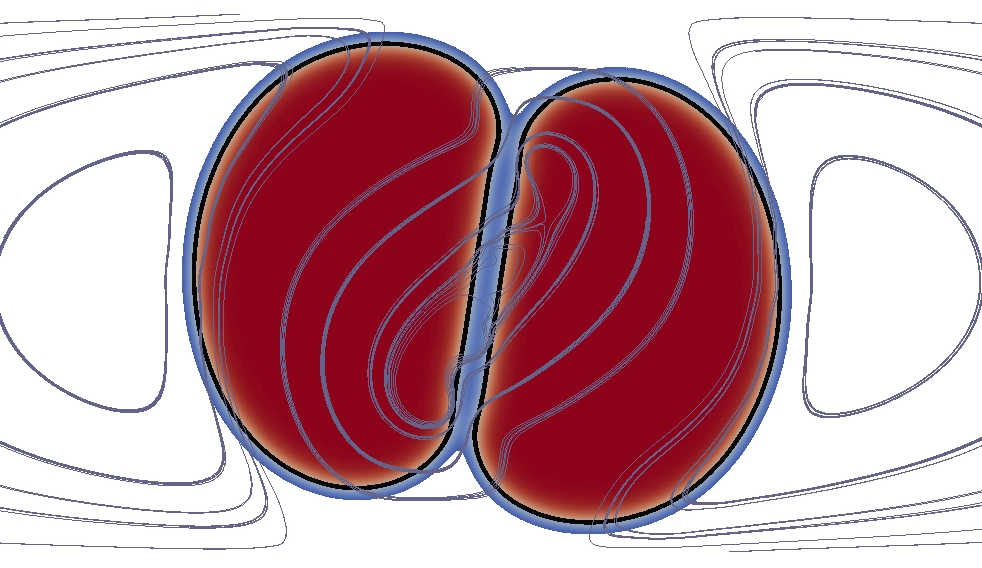}
}&
\subfloat{
\includegraphics[width=0.21\textwidth]{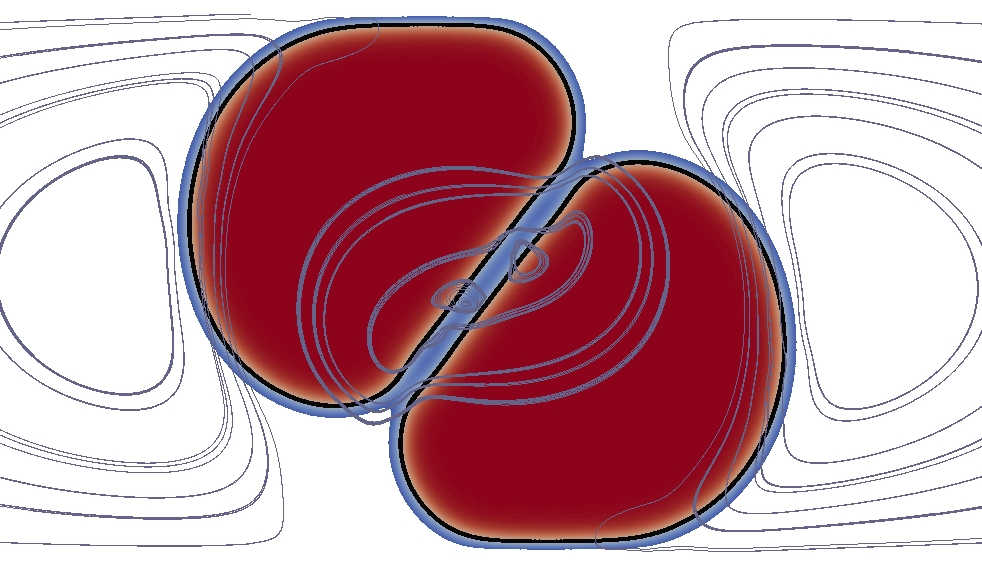}
}
&
\subfloat{
\includegraphics[width=0.21\textwidth]{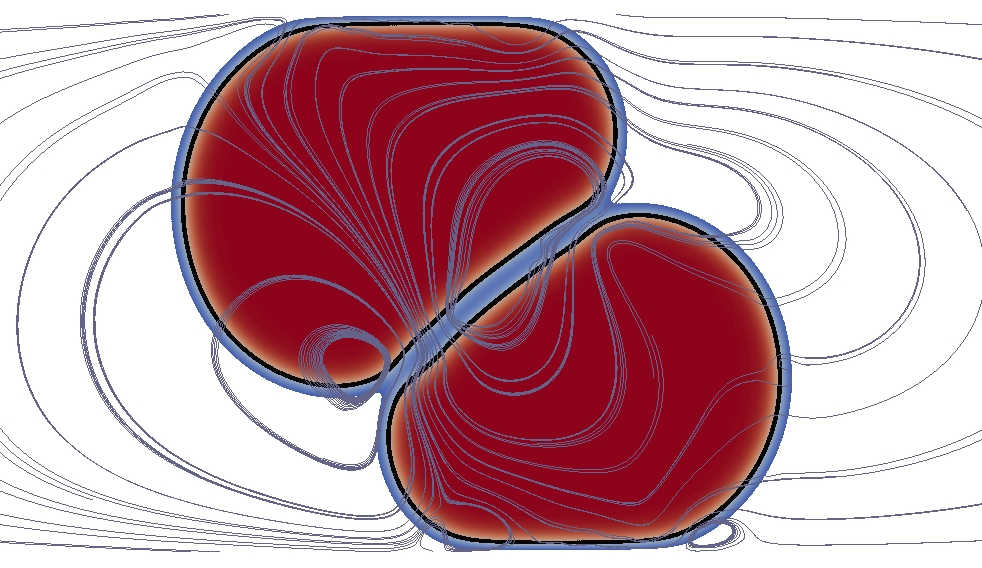}
}
\\
\vspace{-2.5ex}
\raisebox{5ex}{\footnotesize\bf{(b)}} &
\subfloat{
\includegraphics[width=0.21\textwidth]{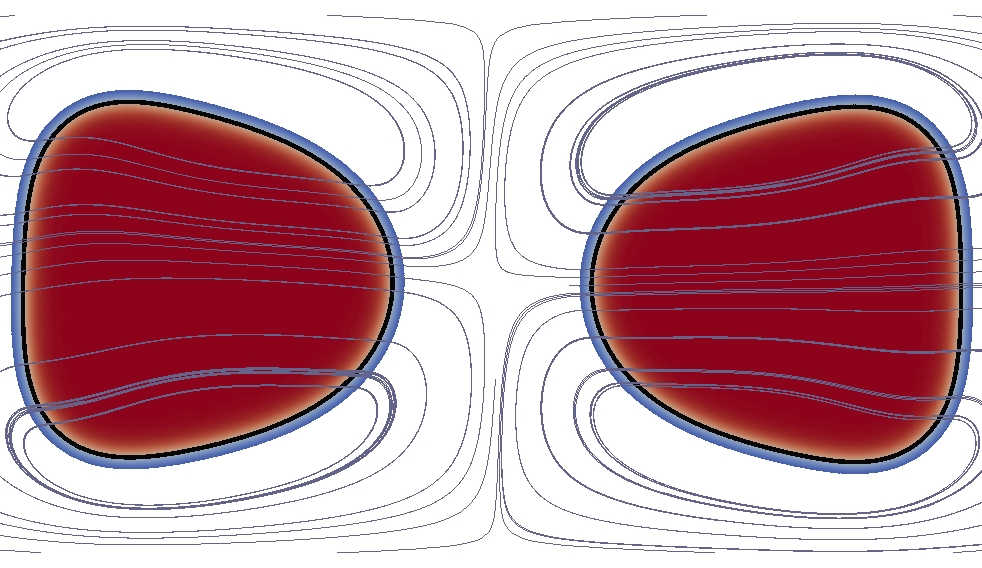}
}&
\subfloat{
\includegraphics[width=0.21\textwidth]{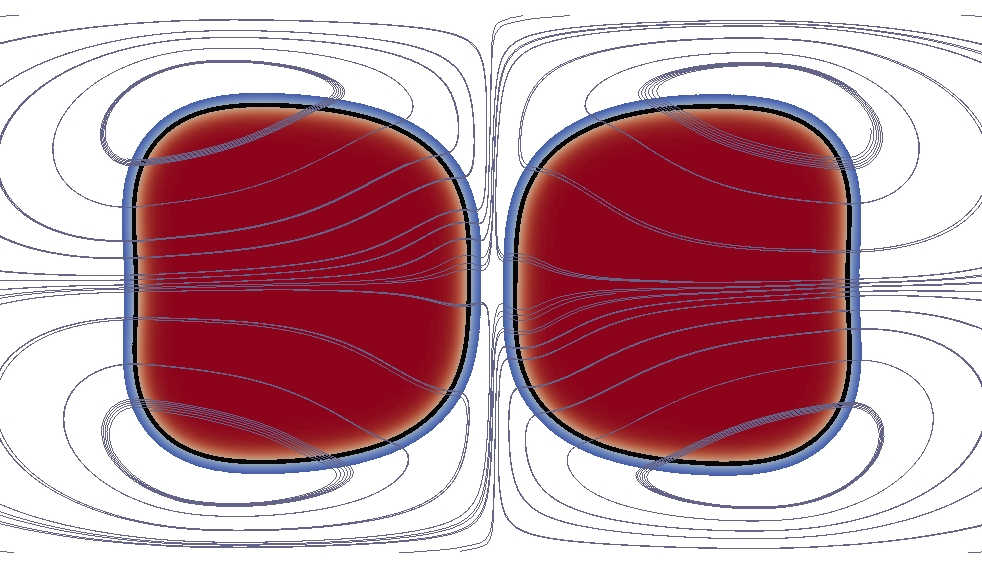}
}&
\subfloat{
\includegraphics[width=0.21\textwidth]{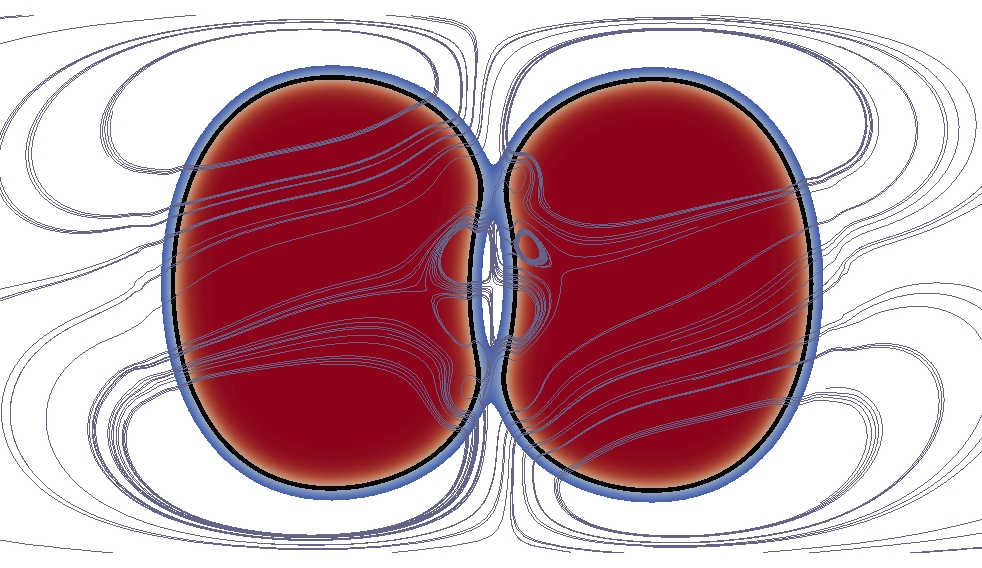}
}
&
\subfloat{
\includegraphics[width=0.21\textwidth]{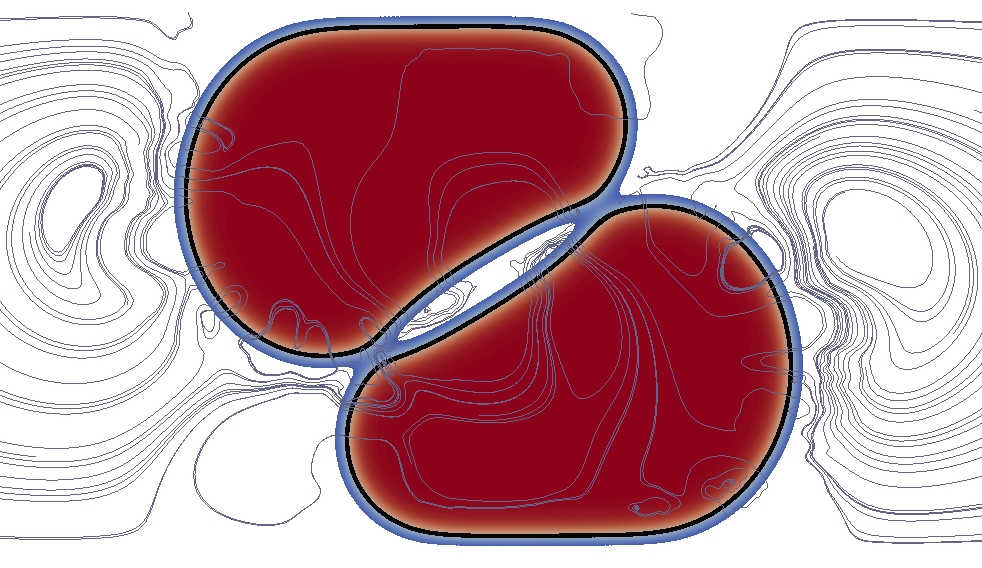}
}
\\
\vspace{-2.5ex}
\raisebox{5ex}{\footnotesize\bf{(c)}} & 
\subfloat{
\includegraphics[width=0.21\textwidth]{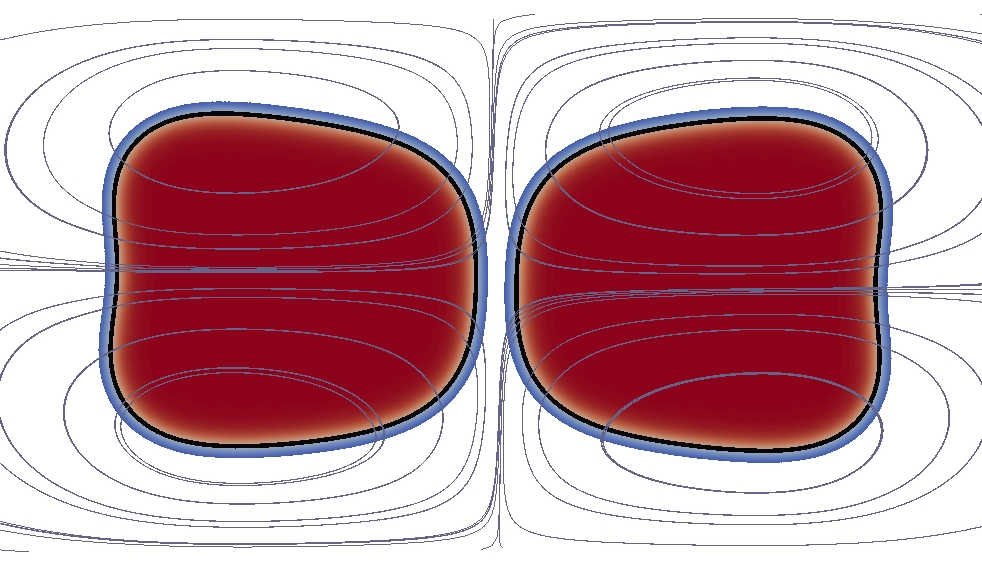}
}&
\subfloat{
\includegraphics[width=0.21\textwidth]{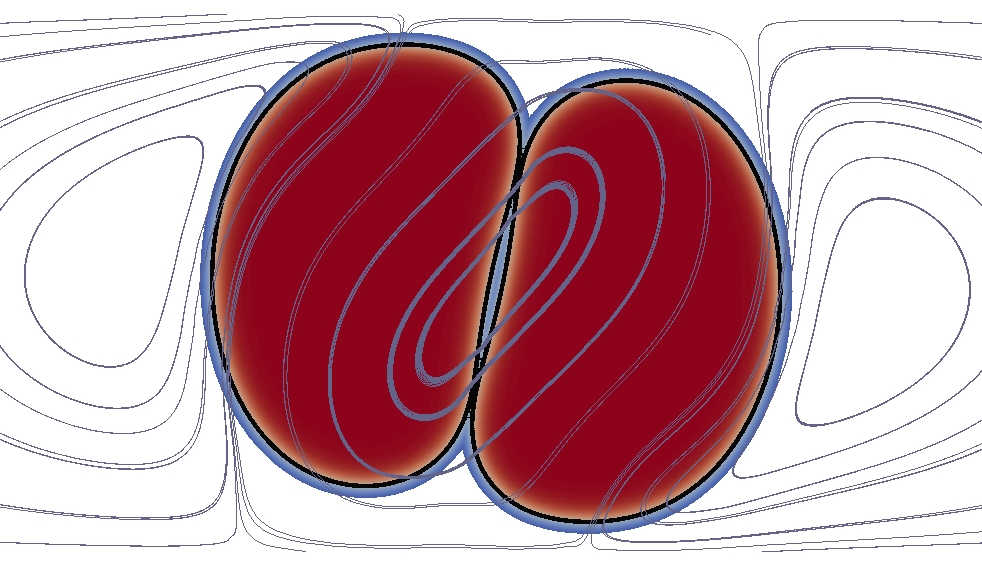}
}&
\subfloat{
\includegraphics[width=0.21\textwidth]{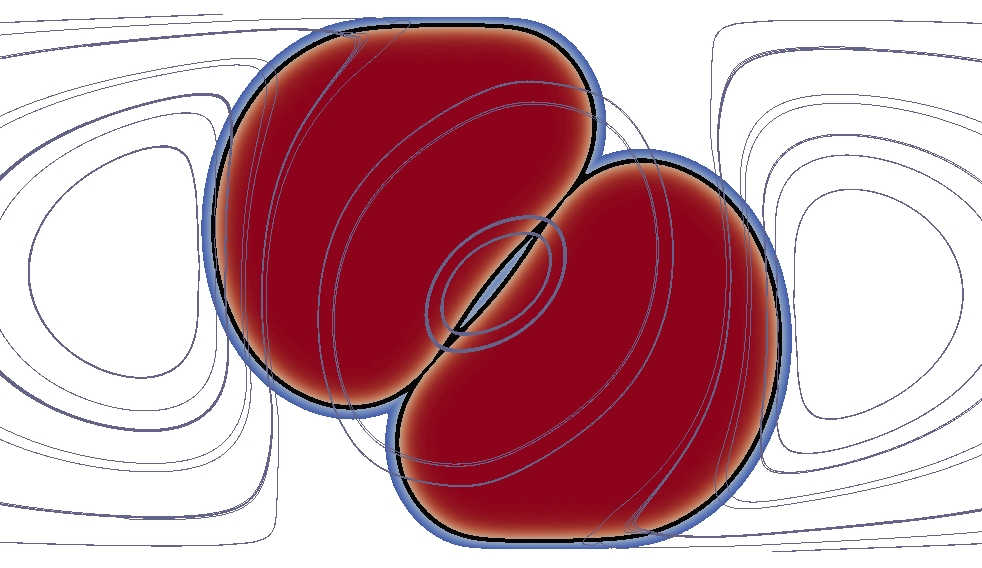}
}
&
\subfloat{
\includegraphics[width=0.21\textwidth]{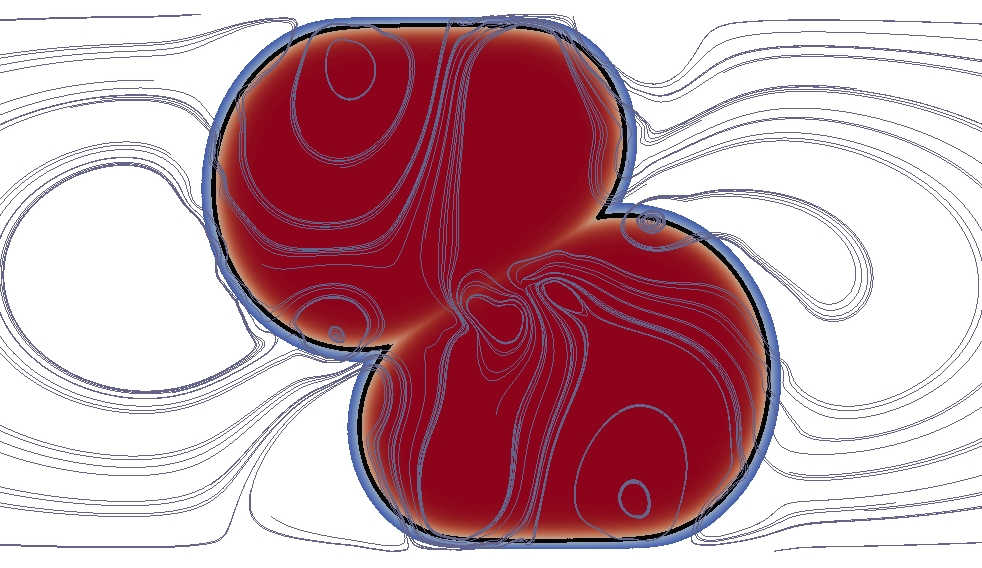}
}
\\
\vspace{-2.5ex}
\raisebox{5ex}{\footnotesize\bf{(d)}} & 
\subfloat{
\includegraphics[width=0.21\textwidth]{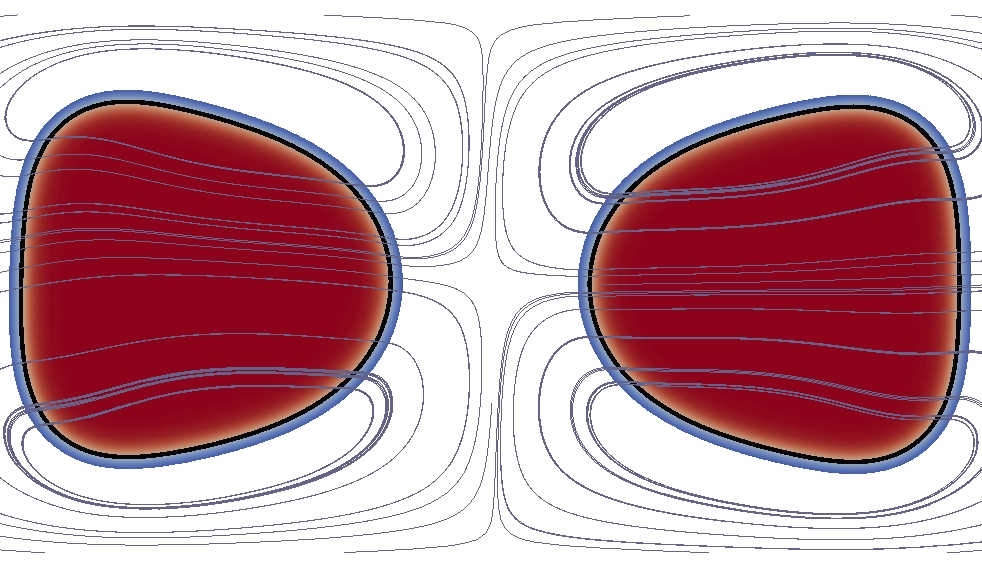}
}&
\subfloat{
\includegraphics[width=0.21\textwidth]{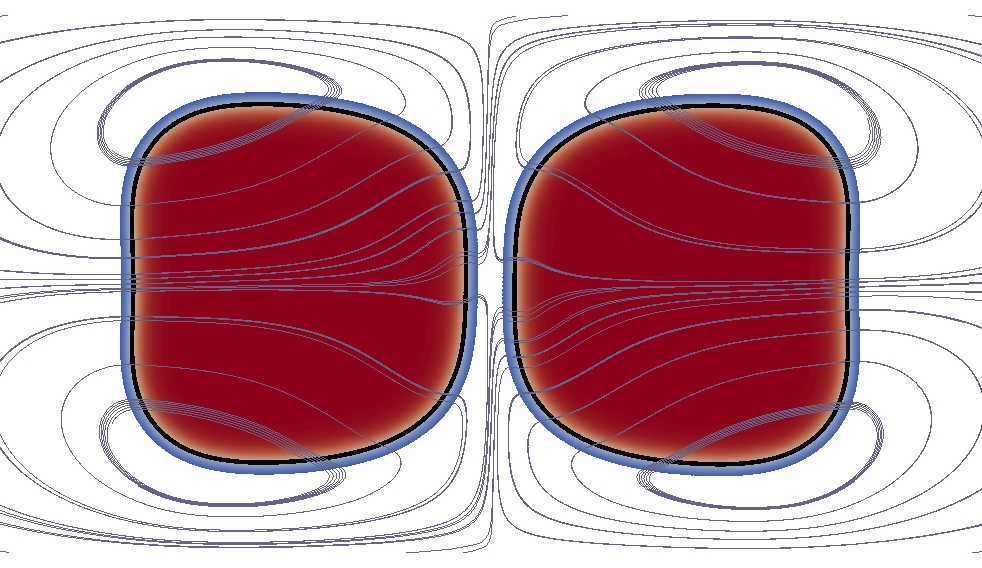}
}&
\subfloat{
\includegraphics[width=0.21\textwidth]{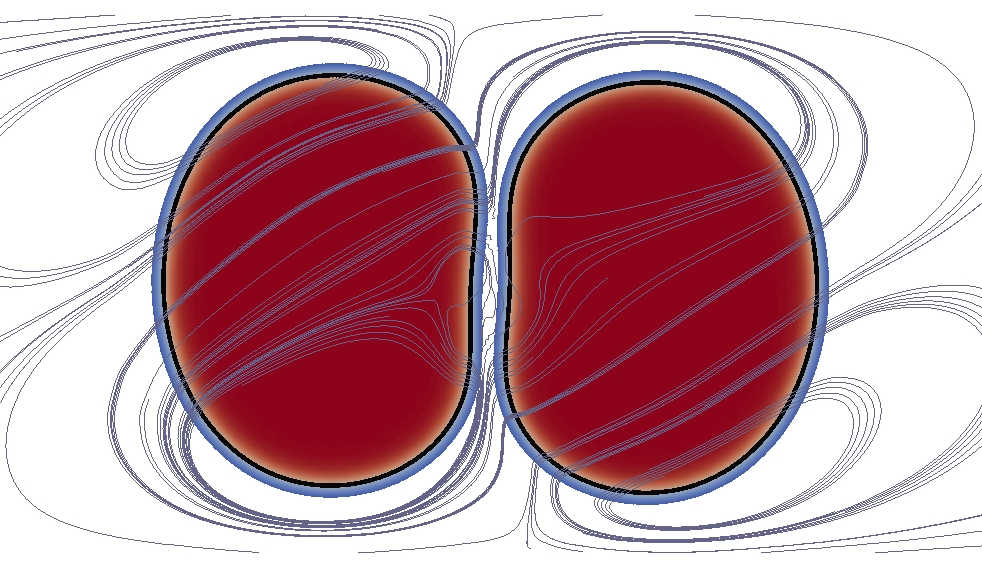}
}
&
\subfloat{
\includegraphics[width=0.21\textwidth]{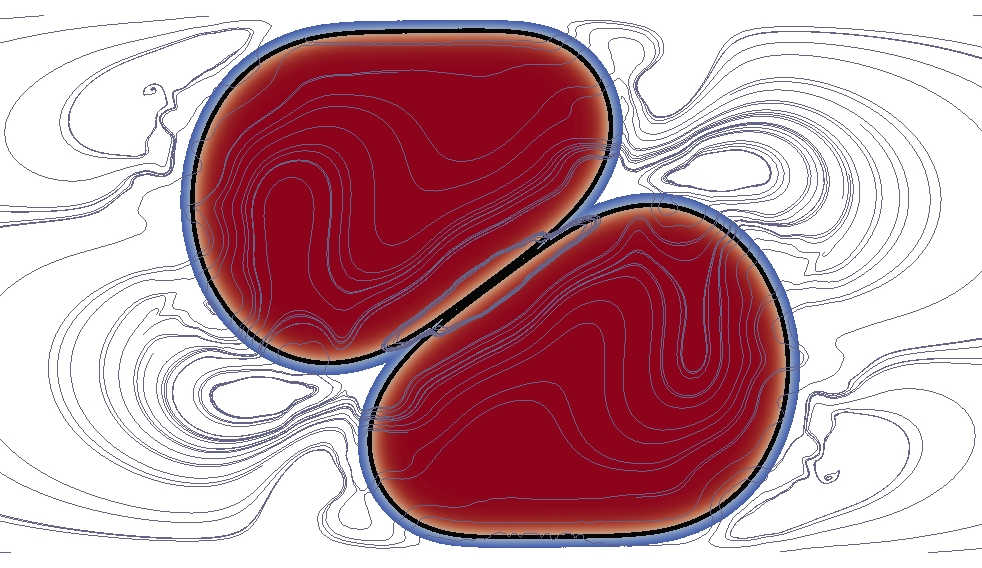}
}
\\
\vspace{-2.5ex}
\raisebox{5ex}{\footnotesize\bf{(e)}} & 
\subfloat{
\includegraphics[width=0.21\textwidth]{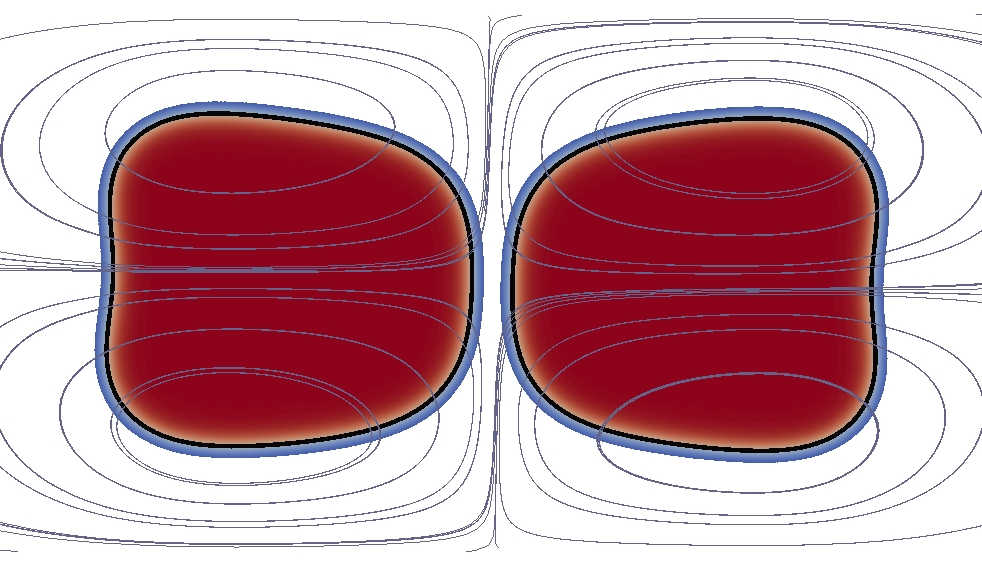}
}&
\subfloat{
\includegraphics[width=0.21\textwidth]{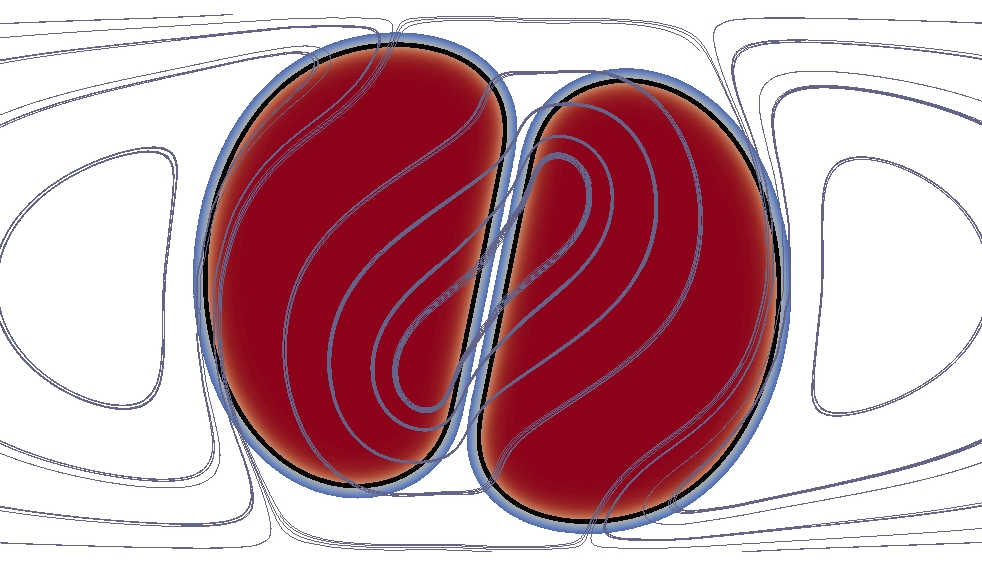}
}&
\subfloat{
\includegraphics[width=0.21\textwidth]{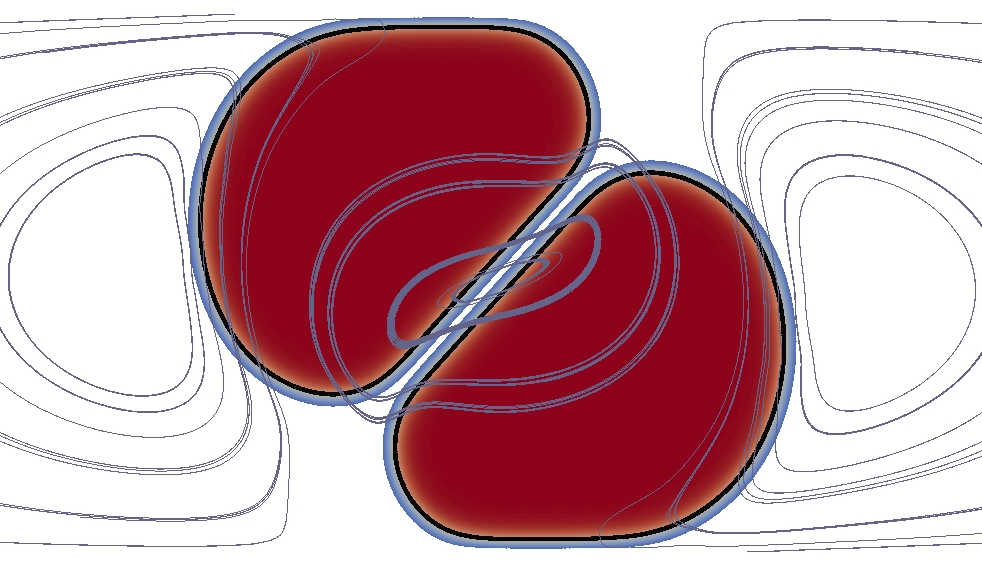}
}
&
\subfloat{
\includegraphics[width=0.21\textwidth]{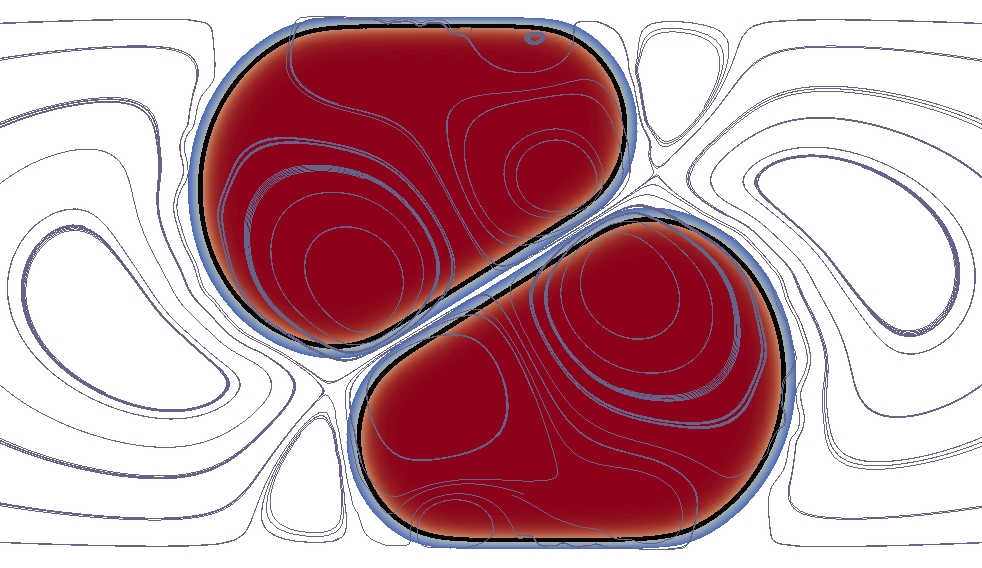}
}
\\
\raisebox{5ex}{\footnotesize\bf{(f)}} & 
\subfloat{
\includegraphics[width=0.21\textwidth]{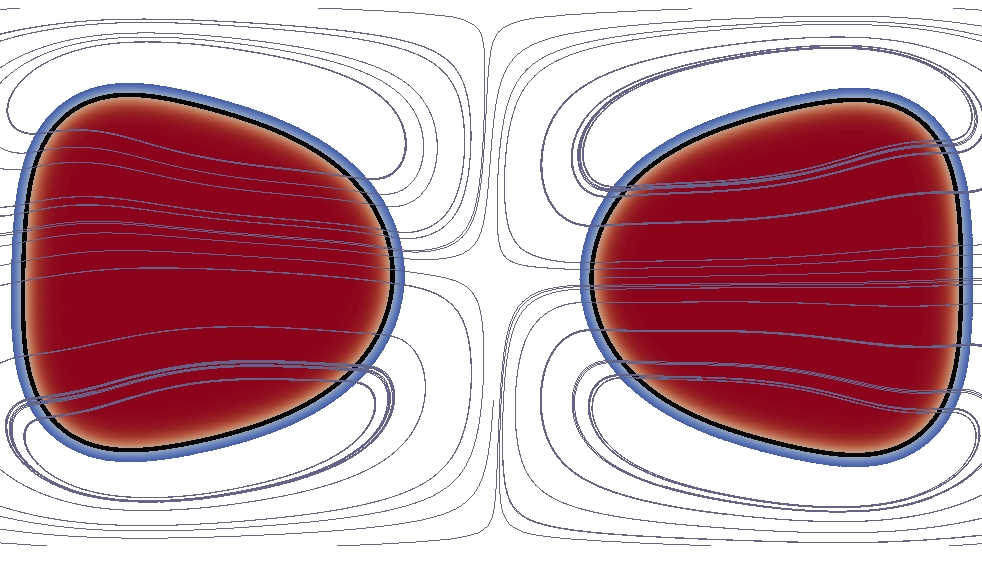}
}&
\subfloat{
\includegraphics[width=0.21\textwidth]{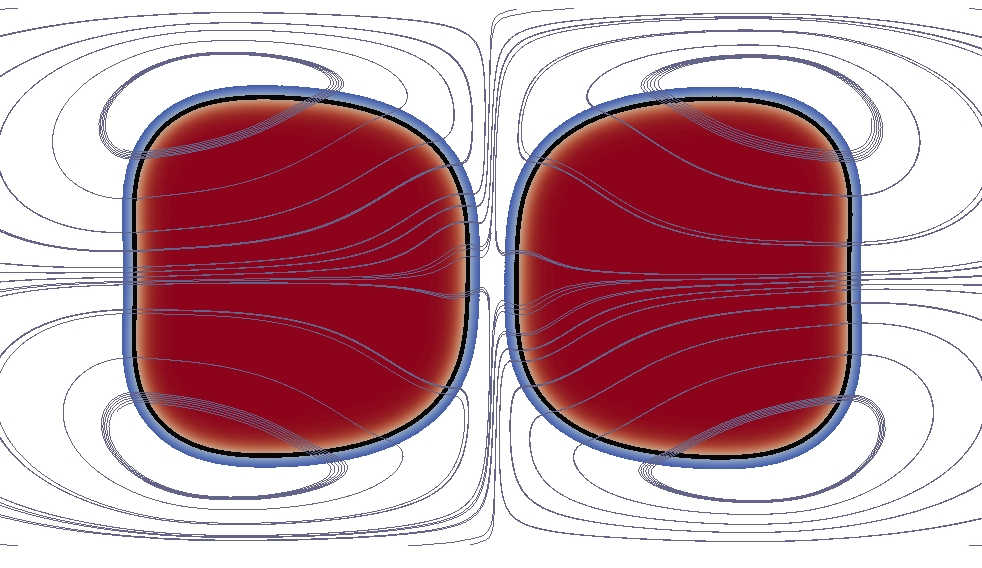}
}&
\subfloat{
\includegraphics[width=0.21\textwidth]{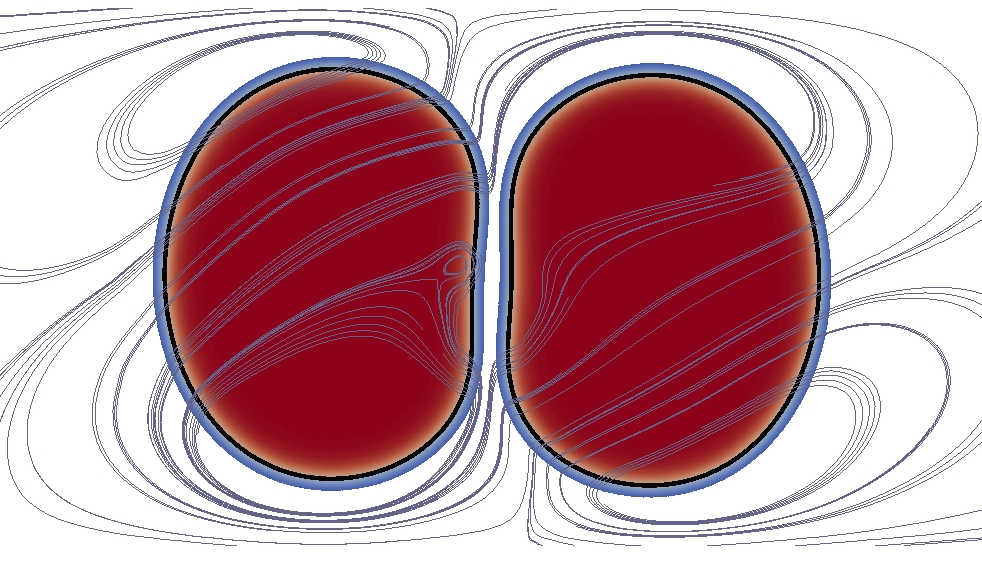}
}
&
\subfloat{
\includegraphics[width=0.21\textwidth]{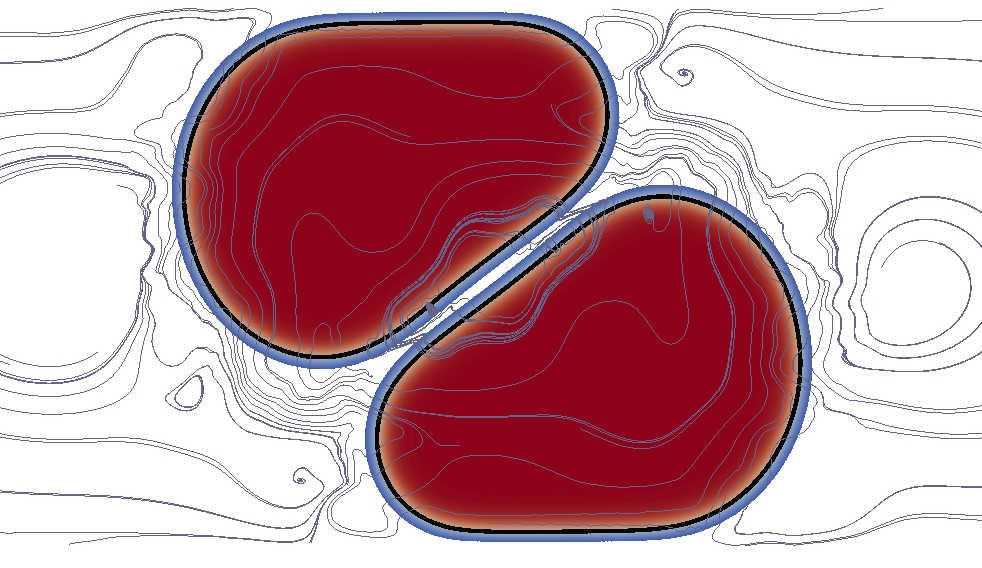}
}
\\
\end{tabular}
\caption{snapshots of the cell collision for time $0.2$, $0.4$, $0.6$ and $2.5$ (from left to right). The rows indicate the used approaches: (a) one phase field, (b) one phase field with inextensibility, (c) two phase fields, (d) two phase fields with inextensibility, (e) two phase fields with interaction potential (Id$=0.01$) and finally (f) two phase fields with interaction potential (Id$=0.01$) and inextensibility. Shown are the streamlines of the velocity field as well as the $[-0.8,1]$ level set of the $\phi_{cell}$ and the zero level set (black line)}
\label{fig:collision}
\end{figure}

We set up two elliptical cells at $(1.45,0.88)$ and $(3.8, 0.875)$, with axis length $0.5\sqrt{2}$ in $x_2$-direction and $0.5$ in $x_1$-direction, where the first cell is placed a little bit higher in order to have comparable situations at coalescence, with the left cell always on top. The computational domain is $[0, 5.25]\times[0,1.75]$. We apply Dirichlet conditions on each boundary. To provide a collision, we adopt a space dependent volume force $\mathbf{F}=(0.5(\phi_{cell}+1) \frac{1}{\text{Fr}} f_1, 0)^\top$ added to the right hand side of eq. \ref{eq:navierStokesNonD1}, with $f_1=1$, if  $x_1<2.625$ and $f_1=-1$ if $x_1>2.625$, and choose the Froude number Fr=$10^{-5}$. We set Re$=0.01$, Be$=5$, mobility $\gamma=10^{-5}$ and a viscosity ratio $\nu_{cell}/\nu_0=10$. The evolution of the cells at time 0.2, 0.4, 0.6 and 2.5 is shown in Fig. \ref{fig:collision}. Each row considers a different modeling approach. We consider six cases: one phase field without (a) and with (b) inextensibility constraint, two phase fields without (c) and with (d) inextensibility constraint and two phase field and an interaction potential without (e) and with (f) inextensibility constraint. The last situation describes the used model in the paper. The black lines show the zero level set and the outer interface, e.g. the $[-0.8,0]$ level sets, are colored blue in order to visualize the effects at coalescence.

The simulations show a strong influence of the inextensibility constraint on the dynamics, the cells move more slowly. However, the strongest effect can be seen on coalescence. For one phase field variable the cells come into contact, but do not merge, as merging would cost energy, due to a violation of the $\tanh$ profile. With the inextensibility constraint, which should inhibit the cells from touching according to \cite{Salacetal_JCP_2011, Alandetal_JCP_2014}, the cells come into contact only at two points. However, also in this situation the phase field is irreversibly connected. The results of  \cite{Salacetal_JCP_2011, Alandetal_JCP_2014} thus cannot be generalized. However, in our approach the inextensibility condition is only asymptotically fulfilled, which seems to avoid an entire adhesion but cannot prevent the cells from touching. Overall, in our setting a single phase field is not sufficient to simulate more than one cell if contact cannot be avoided. With two phase fields but no interaction potential the situation is similar. Here, merging of cells by definition is not possible and due to the incompressibility of the fluid, overlapping cells should also not occur. However, the simulations show that cells touch each other, adhere and overlap slightly. This can be reduced by using the inextensibility constraint but not be prevented. Thus, this approach is not sufficient as well. Only with an interaction potential, contact or adhesion of cells was not observed, neigther without nor with the inextensibility constraint. Only the remaining distance between the cells differs and is bigger if the inextensibility constraint is considered. 

\begin{figure}[ht!]
\subfloat[]{
\includegraphics[width=0.3\textwidth]{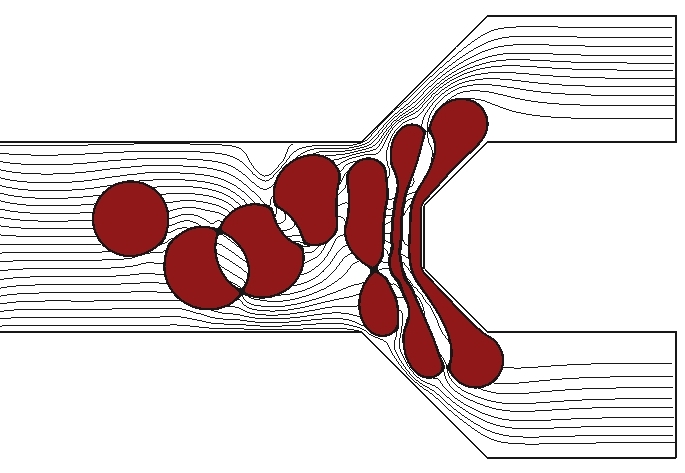}\label{fig:cells_fork1}
}
\subfloat[]{
\includegraphics[width=0.3\textwidth]{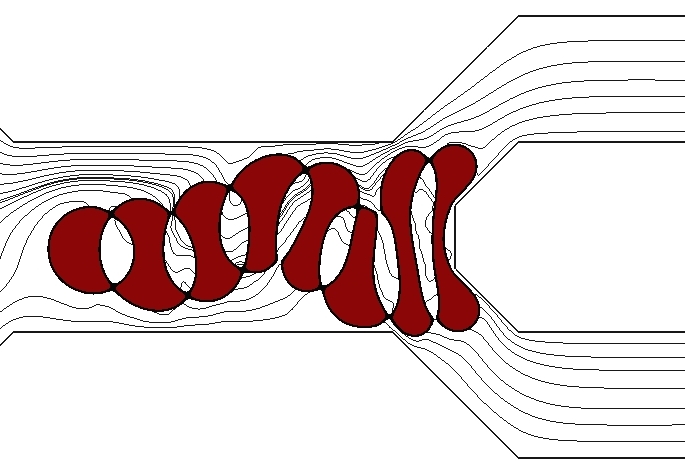}\label{fig:cells_fork2}
}
\subfloat[]{
\includegraphics[width=0.3\textwidth]{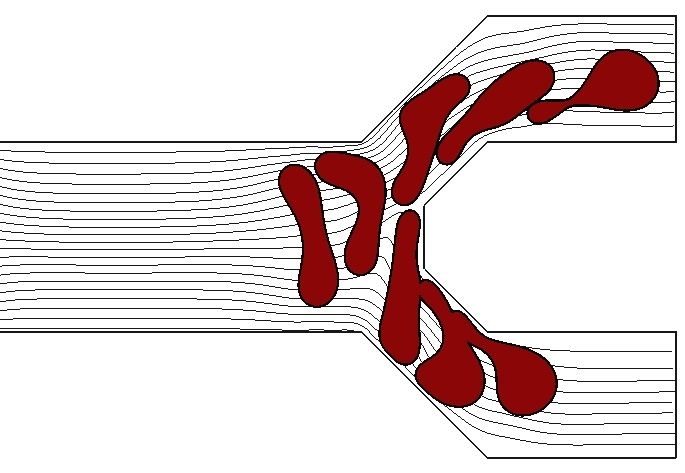}\label{fig:cells_fork3}
}
\vspace{-3ex}
\\
\subfloat[]{
\includegraphics[width=0.3\textwidth]{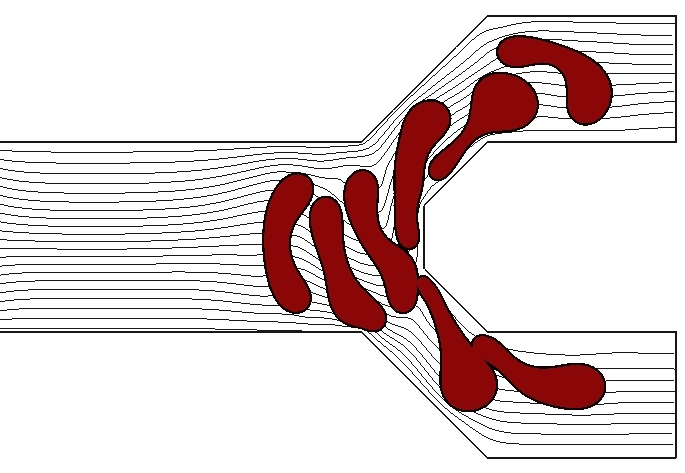}\label{fig:cells_fork4}
}
\subfloat[]{
\includegraphics[width=0.3\textwidth]{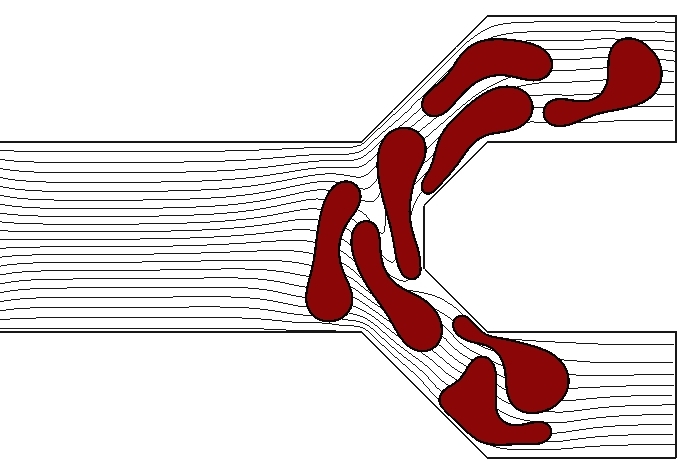}\label{fig:cells_fork5}
}
\subfloat[]{
\includegraphics[width=0.3\textwidth]{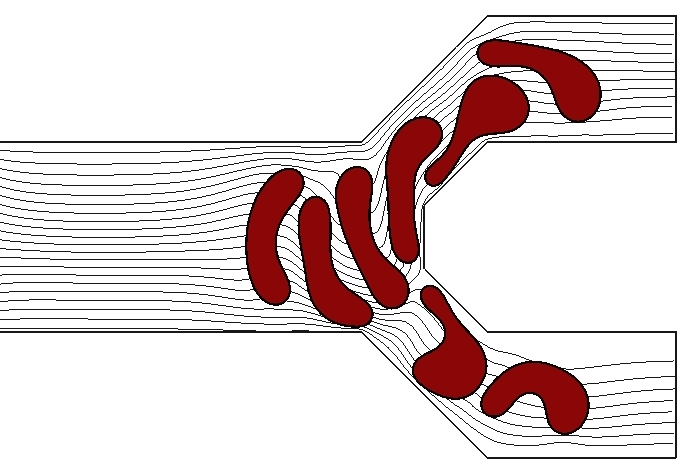}\label{fig:cells_fork6}
}
\caption{Eight RBCs in a symmetric bifurcate vessel (only half of the domain is shown). a) - f) are the same cases as in Fig. \ref{fig:collision}.}
\label{fig:cells_fork}
\end{figure}

\vspace*{0.3cm}
\noindent{\bf Benchmark - RBCs in a bifurcate vessel}\\
In the following, we consider a more realistic setting, where we put eight RBCs in a flow inside a bifurcate vessel. We consider similar parameters as before, Re$=1.125\cdot 10^{-4}$, Be$_{RBC}=2.5$, Id$=0.01$, $\gamma=10^{-7}$ and a viscosity ratio $\nu_{cell}/\nu_0=10$. Flow is considered through the force term $\mathbf{F}=(\frac{1}{\text{Fr}}, 0)^\top$, with Fr$=2.4\cdot 10^{-6}$ leading to a maximal velocity of magnitude 10. Fig. \ref{fig:cells_fork} shows the results for the same eight cases as considered above. Also for this situation only the cases with one phase field for each cell and an interaction potential lead to acceptable results. Differences in the dynamics still can be observed for the case with and without the inextensibility constraint.

\section*{Acknowledgments}

Simulations were carried out at ZIH at TU Dresden and JSC at FZ J\"ulich. W.M. and A.V. acknowledge support from the German Science Foundation through grant no. Vo-899/11. We further acknowledge support from the European Commission within FP7-PEOPLE-2009-IRSES PHASEFIELD and computing resources at JSC through grant no. HDR06.


\bibliographystyle{plos2009.bst}      
\bibliography{literature_wbc_marg}   

\begin{thebibliography}{10}
\providecommand{\url}[1]{\texttt{#1}}
\providecommand{\urlprefix}{URL }
\expandafter\ifx\csname urlstyle\endcsname\relax
  \providecommand{\doi}[1]{doi:\discretionary{}{}{}#1}\else
  \providecommand{\doi}{doi:\discretionary{}{}{}\begingroup
  \urlstyle{rm}\Url}\fi
\providecommand{\bibAnnoteFile}[1]{%
  \IfFileExists{#1}{\begin{quotation}\noindent\textsc{Key:} #1\\
  \textsc{Annotation:}\ \input{#1}\end{quotation}}{}}
\providecommand{\bibAnnote}[2]{%
  \begin{quotation}\noindent\textsc{Key:} #1\\
  \textsc{Annotation:}\ #2\end{quotation}}
\providecommand{\eprint}[2][]{\url{#2}}

\bibitem{Fedosovetal_PRL_2012}
Fedosov DA, Fornleitner J, Gompper G (2012) Margination of white blood cells in
  microcapillary flow.
\newblock Physical Review Letters 108: 1--5.
\bibAnnoteFile{Fedosovetal_PRL_2012}

\bibitem{Fedosovetal_SM_2014}
Fedosov DA, Gompper G (2014) White blood cell margination in microcirculation.
\newblock Soft Matter 10: 2961--2970.
\bibAnnoteFile{Fedosovetal_SM_2014}

\bibitem{Pearsonetal_AJP_2000}
Pearson MJ, Lipowsky HH (2000) Influence of erythrocyte aggregation on
  leukocyte margination in postcapillary venules of rat mesentery.
\newblock American Journal of Physiology - Heart and Circulatory Physiology
  279: H1460--H1471.
\bibAnnoteFile{Pearsonetal_AJP_2000}

\bibitem{Abbittetal_AJP_2003}
Abbitt KB, Nash GB (2003) Rheological properties of the blood influencing
  selectin-mediated adhesion of flowing leukocytes.
\newblock American Journal of Physiology - Heart and Circulatory Physiology
  285: H229--H240.
\bibAnnoteFile{Abbittetal_AJP_2003}

\bibitem{Jainetal_PLOSONE_2009}
Jain A, Munn LL (2009) Determinants of leukocyte margination in rectangular
  microchannels.
\newblock PLoS ONE 4: e7104.
\bibAnnoteFile{Jainetal_PLOSONE_2009}

\bibitem{Takeishi_PhysRep_2014}
Takeishi N, Imai Y, Nakaaki K, Yamaguchi T, Ishikawa T (2014) Leukocyte
  margination at arteriole shear rate.
\newblock Physiological Reports 2: e12037.
\bibAnnoteFile{Takeishi_PhysRep_2014}

\bibitem{Helfrich_ZNF_1973}
Helfrich W (1973) Elastic properties of lipid bilayers: theory and possible
  experiments.
\newblock Zeitschrift fuer Naturforschung, C 28: 693--703.
\bibAnnoteFile{Helfrich_ZNF_1973}

\bibitem{Freund_PF_2007}
Freund JB (2007) Leukocyte margination in a model microvessel.
\newblock Physics of Fluids 19: 023301.
\bibAnnoteFile{Freund_PF_2007}

\bibitem{Barketal_JBM_2010}
Bark DL, Ku DN (2010) Wall shear over high degree stenoses pertinent to
  atherothrombosis.
\newblock Journal of Biomechanics 43: 2970--2977.
\bibAnnoteFile{Barketal_JBM_2010}

\bibitem{Vennemannetal_EF_2007}
Vennemann P, Lindken R, Westerweel J (2007) In vivo whole-field blood velocity
  measurement techniques.
\newblock Experiments in Fluids 42: 495--511.
\bibAnnoteFile{Vennemannetal_EF_2007}

\bibitem{Fischeretal_Science_1978}
Fischer TM, St\"{o}hr-Liesen M, Schmid-Sch\"{o}nbein H (1978) The red-cell as a
  fluid droplet: tank tread-like motion of the human erythocyte-membrane in
  shear flow.
\newblock Science 202: 894--896.
\bibAnnoteFile{Fischeretal_Science_1978}

\bibitem{Krausetal_PRL_1996}
Kraus M, Wintz W, Seifert U, Lipowsky R (1996) Fluid vesicle in shear flow.
\newblock Physical Review Letters 77: 3685--3688.
\bibAnnoteFile{Krausetal_PRL_1996}

\bibitem{Bibenetal_PRE_2003}
Biben T, Misbah C (2003) Tumbling of vesicles under shear flow within an
  advected-field approach.
\newblock Physical Review E 67: 031908.
\bibAnnoteFile{Bibenetal_PRE_2003}

\bibitem{Beaucourtetal_PRE_2004}
Beaucourt J, Rioual F, Seon T, Biben T, Misbah C (2004) Steady to unsteady
  dynamics of a vesicle in a flow.
\newblock Physical Review E 69: 011906.
\bibAnnoteFile{Beaucourtetal_PRE_2004}

\bibitem{Bibenetal_PRE_2005}
Biben T, Kassner K, Misbah C (2005) Phase-field approach to three-dimensional
  vesicle dynamics.
\newblock Physical Review E 72: 041921.
\bibAnnoteFile{Bibenetal_PRE_2005}

\bibitem{Veerapanemietal_JCP_2009}
Veerapaneni S, Gueyffier D, Zorin D, Biros G (2009) A boundary integral method
  for simulating the dynamics of inextensible vesicles suspended in a viscous
  fluid in 2d.
\newblock Journal of Computational Physics 228: 2334--2353.
\bibAnnoteFile{Veerapanemietal_JCP_2009}

\bibitem{Ghigliottietal_JFM_2010}
Ghigliotti G, Biben T, Misbah C (2010) Rheology of a dilute two-dimensional
  suspension of vesicles.
\newblock Journal of Fluid Mechanics 653: 489--518.
\bibAnnoteFile{Ghigliottietal_JFM_2010}

\bibitem{Sohnetal_JCP_2010}
Sohn JS, Tseng YH, Li S, Voigt A, Lowengrub J (2010) Dynamics of multicomponent
  vesicles in a viscous fluid.
\newblock Journal of Computational Physics 229: 119--144.
\bibAnnoteFile{Sohnetal_JCP_2010}

\bibitem{Kimetal_JCP_2010}
Kim Y, Lai MC (2010) Simulating the dynamics of inextensible vesicles by the
  penalty immersed boundary method.
\newblock Journal of Computational Physics 229: 4840--4853.
\bibAnnoteFile{Kimetal_JCP_2010}

\bibitem{Zhaoetal_JFM_2011}
Zhao H, Shaqfeh ESG (2011) The dynamics of a vesicle in simple shear flow.
\newblock Journal of Fluid Mechanics 674: 578--604.
\bibAnnoteFile{Zhaoetal_JFM_2011}

\bibitem{Laadharietal_PF_2012}
Laadhari A, Saramito P, Misbah C (2012) Vesicle tumbling inhibited by inertia.
\newblock Physica of Fluids 24: 031901.
\bibAnnoteFile{Laadharietal_PF_2012}

\bibitem{Salacetal_JFM_2012}
Salac D, Miksis MJ (2012) Reynolds number effects on lipid vesicles.
\newblock Journal of Fluid Mechanics 711: 122--146.
\bibAnnoteFile{Salacetal_JFM_2012}

\bibitem{Alandetal_JCP_2014}
Aland S, Egerer S, Lowengrub J, Voigt A (2014) Diffuse interface models of
  locally inextensible vesicles in a viscous fluid.
\newblock Journal of Computational Physics 277: 32--47.
\bibAnnoteFile{Alandetal_JCP_2014}

\bibitem{Bonitoetal_MMNP_2011}
Bonito A, Nochetto RH, Pauletti MS (2011) Dynamics of biomembranes: effect of
  the bulk fluid.
\newblock Mathematical Modelling of Natural Phenomena 6: 25--43.
\bibAnnoteFile{Bonitoetal_MMNP_2011}

\bibitem{Duetal_DCDS_2007}
Du Q, Li M, Liu C (2007) Analysis of a phase field navier-stokes vesicle-fluid
  interaction model.
\newblock Discrete and Continuous Dynamical Systems - Series B 8: 539--556.
\bibAnnoteFile{Duetal_DCDS_2007}

\bibitem{Marthetal_JMB_2013}
Marth W, Voigt A (2014) Signaling networks and cell motility: a computational
  approach using a phase field description.
\newblock Journal of Mathematical Biology 69: 91--112.
\bibAnnoteFile{Marthetal_JMB_2013}

\bibitem{Hausseretal_IJBMBS_2013}
Hau{\ss}er F, Li S, Lowengrub J, Marth W, R\"{a}tz A, et~al. (2013)
  Thermodynamically consistent models for two-component vesicles.
\newblock International Journal of Biomathematics and Biostatistics 2: 19--48.
\bibAnnoteFile{Hausseretal_IJBMBS_2013}

\bibitem{Salacetal_JCP_2011}
Salac D, Miksis M (2011) A level set projection model of lipid vesicles in
  general flows.
\newblock Journal of Computational Physics 230: 8192--8215.
\bibAnnoteFile{Salacetal_JCP_2011}

\bibitem{Duetal_Nonl_2005}
Du Q, Liu C, Ryham R, Wang X (2005) A phase field formulation of the {W}illmore
  problem.
\newblock Nonlinearity 18: 1249-1267.
\bibAnnoteFile{Duetal_Nonl_2005}

\bibitem{Campeloetal_EPJE_2007}
Campelo F, Hern\'{a}ndez-Machado A (2007) Shape instabilities in vesicles: A
  phase-field model.
\newblock The European Physical Journal Special Topics 143: 101--108.
\bibAnnoteFile{Campeloetal_EPJE_2007}

\bibitem{Du_JCP_2006}
Du Q, Liu C, Wang X (2006) Simulating the deformation of vesicle membranes
  under elastic bending energy in three dimensions.
\newblock Journal of Computational Physics 212: 757--777.
\bibAnnoteFile{Du_JCP_2006}

\bibitem{Zhangetal_JCP_2009}
Zhang J, Das S, Du Q (2009) A phase field model for vesicle-substrate adhesion.
\newblock Journal of Computational Physics 228: 7837--7849.
\bibAnnoteFile{Zhangetal_JCP_2009}

\bibitem{Guetal_JCP_2014}
Gu R, Wang X, Gunzburger M (2014) Simulating vesicle-substrate adhesion using
  two phase field functions.
\newblock Journal of Computational Physics 275: 626--641.
\bibAnnoteFile{Guetal_JCP_2014}

\bibitem{Tanaka_PRL_2000}
Tanaka H, Araki T (2000) Simulation method of colloidal suspensions with
  hydrodynamic interactions: Fluid particle dynamics.
\newblock Physical Review Letters 85: 1338--1341.
\bibAnnoteFile{Tanaka_PRL_2000}

\bibitem{Veyetal_CVS_2007}
Vey S, Voigt A (2007) {AMDiS}: adaptive multidimensional simulations.
\newblock Computing and Visualization in Science 10: 57--67.
\bibAnnoteFile{Veyetal_CVS_2007}

\bibitem{Witkowskietal_ACM_2015}
Witkowski T, Ling S, Praetorius S, Voigt A (2015) Software concepts and
  numerical algorithms for a scalable adaptive parallel finite element method.
\newblock Advances in Computational Mathematics : DOI:
  10.1007/s10444-015-9405-4.
\bibAnnoteFile{Witkowskietal_ACM_2015}

\bibitem{Voigtetal_JCS_2012}
Voigt A, Witkowski T (2012) A multi-mesh finite element method for lagrange
  elements of arbitrary degree.
\newblock Journal of Computational Science 3: 420--428.
\bibAnnoteFile{Voigtetal_JCS_2012}

\end{thebibliography}

\end{document}